\documentclass[showpacs,amsmath,amssymb,prb,aps,twocolumn]{revtex4}
\usepackage{graphicx}
\begin{document}
\title {Coulomb corrections to the
extrinsic spin-Hall effect of a two-dimensional electron gas}
\author {E. M. Hankiewicz and G. Vignale}
\email{vignaleg@missouri.edu}
\affiliation{Department of Physics
and Astronomy University of Missouri, Columbia, Missouri 65211,
USA}
\date{\today}
\begin{abstract}
We develop the microscopic theory of the extrinsic spin Hall
conductivity  of a two-dimensional electron gas, including
skew-scattering, side-jump, and Coulomb interaction effects. We find
that while the spin-Hall conductivity connected with the side-jump
is independent of the strength of electron-electron interactions,
the skew-scattering  term is reduced by the spin-Coulomb drag, so
the total spin current and the total spin-Hall conductivity are
reduced for typical experimental mobilities. Further, we predict
that in paramagnetic systems the spin-Coulomb drag reduces the
spin accumulations in two different ways: (i) directly through the
reduction of the skew-scattering contribution (ii)
indirectly through the reduction of the spin diffusion length.
Explicit expressions for the various contributions to the spin Hall conductivity are
obtained using an exactly solvable model of the skew-scattering.
\end{abstract}
 \maketitle
\begin{widetext}
\section {Introduction}
There has recently been a strong revival of interest in the
phenomenon of the {\it spin Hall effect} in the context of
semiconductor spintronics \cite{Wolf01}.  Two different forms of
this phenomenon have been identified:  an extrinsic spin-Hall
effect \cite{Dyakonov71,Perel,Hirsch99,Zhang00}, which is driven
by spin-orbit scattering with impurities, and an intrinsic one
\cite{Murakami03,Sinova04,Murakami05r}, which is due to spin-orbit
effects in the  band structure. In both forms, the phenomenon
consists of the appearance of a transverse {\it spin current} --
say a $z$-spin current in the $y$-direction --   when the electron
gas is driven by an electric field in the $x$-direction.  The
physical manifestations of this spin current are still an active
subject of study and controversy.\cite{Rashba03}  However, it is
now believed that the spin Hall effect should lead to transverse
spin accumulation when the flow of the spin current is suppressed
by an appropriate gradient of spin-dependent electrochemical
potential. Recently, the spin Hall effect has been experimentally
observed in Kerr rotation experiments in 3D and 2D n-doped
GaAs\cite{Kato04,Sih05} and in a p-n junction light emitting diode
(LED) realized in a two-dimensional hole system
\cite{Wunderlich05}. However, there is still a debate on the
origin (extrinsic or intrinsic) of  the experimentally observed
spin
accumulation.\cite{Engel05,Bernevig04,Nikolic04,Sarma05,Sarma05a}

In this paper we focus exclusively on the theory of the extrinsic
effect in a two-dimensional electron gas (2DEG). It has long been
realized that the extrinsic spin Hall current is the sum of two
contributions\cite{Nozieres}. The first contribution (commonly
known as ``skew-scattering" mechanism \cite{Smit55,Smit58}) arises
from the asymmetry of the electron-impurity scattering in the
presence of spin-orbit interactions \cite{Mott}: electrons that
are drifting in the $+x$ direction under the action of an electric
field are more likely to be scattered to the left than to the
right if, say,  their spin is up, while the reverse is true if
their spin is down. This generates a net $z$-spin current in the
$y$ direction.   The second contribution (the so-called
``side-jump" mechanism \cite{Berger70a,Berger70b,Lyo72,Nozieres})
is caused by the anomalous  relationship between the physical and
the canonical position operator  (see Eq.~(\ref{r_phys}) below).
This again leads to a finite spin current in the $y$ direction.
The skew-scattering and side-jump contributions were widely
discussed in the context of the anomalous Hall effect
\cite{Luttinger54,Smit55,Smit58,Berger70a,Berger70b,Lyo72,Nozieres,Dugaev01}
in magnetic materials. The skew-scattering contribution was first
studied by Smit \cite{Smit55,Smit58} while the side-jump
contribution was introduced by Berger\cite{Berger70a,Berger70b}.
The theory for both effects has been also discussed recently in
several excellent papers, both for the
extrinsic\cite{Bruno01,Engel05,Sarma05} and the intrinsic
case.\cite{Jungwirth02,Onoda02,Bruno05,Sinitsyn05}

In this paper in addition to the previously considered
skew-scattering and side-jump contributions, we also include the
Coulomb interaction effects. The main effect of interactions on
the spin transport originates from the friction between spin-up
and spin-down electrons moving with different drift velocities,
the so called spin-Coulomb drag (SCD) effect
\cite{Amico00,Flensberg01,Amico02}.  We show that while the
spin-Hall conductivity associated with the side-jump term is
independent of the strength of electron-electron interactions, the
skew scattering part is reduced by the spin-Coulomb drag, so the
absolute value of the spin Hall conductivity (and hence the spin
Hall current) is reduced for experimentally accessible parameters.
 Since the SCD has been predicted (and recently
observed \cite{Orenstein05}) to be a rather significant
contribution to the overall resistivity in high mobility
electronic systems, we think it is important to include it in the
description of the spin Hall effect,  and we show here how this is
done. Moreover, we predict that SCD in paramagnetic materials will
reduce the spin accumulations through the reduction of the
skew-scattering resistivity as well as the spin diffusion length.
Also, we present in the Appendix a simple model for
electron-impurity scattering which can be solved exactly, leading
to an analytical determination of scattering rates as well as
side-jump and skew-scattering contributions to the spin Hall
conductivity.

This paper is organized as follows: in Section II the Hamiltonian
and the Boltzmann equation are presented;  in Section III the
skew-scattering contribution to the conductivity is derived; in
Section IV  we use a force-balance argument to calculate the
side-jump contribution; in Section V the contributions of spin
Coulomb drag and spin-flip processes  are included; in Section VI
the spin accumulation in the presence of skew-scattering,
side-jump, and electron-electron interactions is calculated.
   We summarize the paper in Section VII.
\section {Hamiltonian and Boltzmann
equation} We consider a strictly two-dimensional electron gas
(2DEG)  that  lies in the $x-y$ plane.  The hamiltonian  is
\begin{equation} \label{H}
H=H_0+H_{so}+H_c+H_E~,
\end{equation}
where
\begin{equation}\label{H0}
H_0= \sum_i \left [\frac{p_i^2}{2m} +V_{ei}(\vec r_i)\right]
\end{equation}
 is the non-interacting hamiltonian ($m$ being the effective mass of the conduction band), including the electric
potential $V_{ei}(\vec r)$ generated by randomly distributed
impurities,
\begin{equation}\label{HC} H_c=\frac{1}{2} \sum_{i \neq
j}\frac{e^2}{\epsilon_b|\vec r_i-\vec r_j|}~,
\end{equation}
is the electron-electron interaction (screened by the background
dielectric constant $\epsilon_b$),
\begin{equation}\label{HSO}
H_{so} = \alpha \sum_i \left \{\vec p_i \times \left[\vec \nabla_i
V_{ei}(\vec r_i) + \vec{\nabla}_i V^i_{ee} \right]\right \}\cdot \vec \sigma_i~,
\end{equation}
is the spin-orbit interaction (SOI) induced by the electric
potential of the impurities, $V_{ei}(\vec r_i )$, and of  the
other electrons $\vec V^i_{ee}=\sum_{j \neq
i}\frac{e^2}{\epsilon_b |{\vec r}_i-{\vec r}_j|}$, and finally
\begin{equation}\label{HE}
H_E=\sum_i \left \{ e \vec E \cdot \vec r_i+e \alpha  (\vec p_i \times \vec E) \cdot \vec \sigma_i\right\}
\end{equation}
is the interaction with the external electric field $\vec E$.

The various spin-orbit terms appearing in the hamiltonian can all
be shown to arise from a single basic fact, namely, the change in
form of the physical position operator under  a transformation
that eliminates the coupling between the conduction band in which
the electrons of interest reside, and the spin-orbit-split valence
band. If we denote by $\vec r_i$ the canonical position operator
of the i-th electron, then the physical position operator is given
by
\begin{equation}\label{r_phys}
\vec r_{phys,i} = \vec r_i-\alpha(\vec p_i \times \vec \sigma_i)~,
\end{equation}
and correspondingly the velocity operator is:
\begin{eqnarray}\label{v_phys}
\vec v_i&=& -\frac{i}{\hbar}[\vec r_{phys},\hat H] \nonumber\\ &=&
\frac{\vec p_i}{m}+ 2\alpha\left[\vec \nabla_i V_{ei}({\vec r}_i)
+ \vec \nabla_i V^i_{ee} + e \vec E \right] \times \vec \sigma_i
~,
\end{eqnarray}

The spin-orbit ``coupling constant" $\alpha$ takes into account
the effective SOI induced by the valence bands (heavy holes, light
holes, and split-off band) on conduction electrons in the
framework of the 8-band Kane model. Within this model one
finds\cite{Winkler2003} $$\alpha=\frac{ \hbar
P^2}{3m_e^2}\left[\frac{1}{E_g^2}-\frac{1}{(E_g+\Delta_{SO})^2}\right]~,$$
where $E_g$ is the gap energy between conduction and heavy/light
holes bands, $\Delta_{SO}$ is the splitting energy between
heavy/light holes and split-off bands, $P$ is the matrix element
of the momentum operator between the conduction and the
valence-band edges, and $m_e$ is the bare electron mass. Using
values of the parameters appropriate for the 2DEG in
Al$_{0.1}$Ga$_{0.9}$As \cite{Sih05} with a band mass $m=0.074m_e$
we find: $\alpha\hbar = 4.4\AA^2$ . In this paper we treat the
spin-orbit interaction to the first order in $\alpha$, which is
justified by the smallness of  the parameter $\alpha \hbar/a_B^2$,
where $a_B \simeq 100 \AA$ is the effective Bohr radius. Also, we
consider the first order corrections from the electric potential
associated with electron-electron interactions to the spin-orbit
hamiltonian, i.e. the terms $(\alpha \hbar/a_B^2)(e^2/\epsilon_b
\hbar v_F)$.

Notice that the canonical positions $\vec r_i$ and the canonical momenta
$\vec p_i$ of the particles are vectors in the $x-y$ plane, and so
is the $\vec \nabla$ operator. Therefore $\vec p \times [\vec
\nabla_i V_{ei}(\vec{r_i})+ \vec{\nabla}_i V^i_{ee})] $ is a
vector in the $z$-direction, and the spin-orbit interaction
conserves the $z$-component of the spin of each electron.  This
nice feature of our strictly 2D model allows a particularly simple
analysis of the spin Hall effect without sacrificing any essential
features of the spin-orbit interaction. Processes that flip the
$z$-component of the spin will be considered separately (see
Section V).

We begin to exploit the conservation of $\sigma_z$ by
defining the quasi-classical one-particle distribution function
$f_{\sigma}(\vec r,\vec k,t)$,  i.e. the probability of finding an
electron with $z$-component of the spin $S_z =
\frac{\hbar}{2}\sigma$,  with $\sigma = \pm 1$, at position $\vec r$
with momentum $\vec p = \hbar \vec k$ at the time $t$.   In this
paper we focus on spatially homogeneous steady-state situations, in
which $f_\sigma$ does not depend on $\vec r$ and $t$   (for a
discussion of non-homogeneous spin accumulation effects see Section
VI).  We write
\begin{equation}\label{fsigma}
f_{\sigma}(\vec r,\vec
k,t) =  f_{0\sigma}(\epsilon_k)+f_{1\sigma}(\vec
k)~,
\end{equation}
where $f_{0\sigma}(\epsilon_k)$ is the
equilibrium distribution function -- a function of the free particle
energy $\epsilon_k =\frac{\hbar^2 k^2}{2 m}$ --  and
$f_{1\sigma}(\vec k)$ is a small deviation from equilibrium induced
by the application of  steady electric fields $\vec E_\sigma$
($\sigma=\pm 1$) which couple independently  to each of the two spin
components.  Then to first order in $\vec E_\sigma$ the Boltzmann
equation takes the form
\begin{equation}\label{Boltzmann.Equation}
-e \vec E_\sigma \cdot \frac{\hbar \vec k}{m}
f^\prime_{0\sigma}(\epsilon_k) =\dot f_{1\sigma}(\vec k)_{c}~,
\end{equation}
where $\dot f_{1\sigma}(\vec k)_{c}$   is the first-order  in $\vec
E_\sigma$ part of the collisional time
derivative $\dot f_{\sigma}(\vec k)_{c}$ due to different scattering
processes such as
electron-impurity scattering, electron-electron scattering, and
spin-flip scattering.   As usual, $\dot f_{\sigma}(\vec k)_{c}$
is written  as the difference of an in-scattering and and an
out-scattering terms.  For example, in the case of spin-conserving
electron- impurity scattering one has:
\begin{eqnarray}\label{collision.integral}
\dot f_{\sigma}(\vec k)_{c,imp}  = -\sum_{\vec k'}\left[W_{\vec
k\vec k'\sigma}f_\sigma(\vec k) -W_{\vec k'\vec k\sigma}
f_\sigma(\vec k')\right] \delta(\widetilde
\epsilon_{k\sigma}-\widetilde \epsilon_{k'\sigma})~,
    \end{eqnarray}
where $W_{\vec k\vec k'\sigma}$ is the scattering rate for a
spin-$\sigma$ electron to go from $\vec k$ to $\vec k'$, and
$\widetilde \epsilon_{k \sigma}$ is the particle energy,
including the additional spin-orbit interaction due the electric field $\vec E_\sigma$.

The last point is absolutely vital for a correct accounting of the ``side-jump" contribution.  We must use
\begin{equation}\label{epsilon_tylda}
\widetilde{\epsilon}_{k\sigma}= \epsilon_k +2e\alpha\hbar\sigma (\vec{E}_{\sigma}\times\hat{z})\cdot \vec{k}~,
\end{equation}
where the second term on the right hand side differs by a factor $2$ from what one would surmise
 from the intuitive expression $\epsilon_p+e {\bf E} \cdot {\bf r}_{phys}$.
 Why?  The reason is that the $\delta$-function in Eq.~(\ref{collision.integral}) expresses the
 conservation of energy in a scattering process.  This is a time-dependent process:
  therefore the correct expression  for the change in position of the electron $\Delta {\bf r}_{phys}$
    must be calculated as the integral of the {\it velocity} over time.
    From  the commutator of the physical position operator with the Hamiltonian we easily find
\begin{eqnarray}\label{v_sigma}
\vec v_{i\sigma}&=&  \frac{\vec p_i}{m}+ 2\alpha\sigma\left[\vec
\nabla_i V_{ei}({\vec r}_i) + \vec \nabla_i V^i_{ee} + e \vec E
\right] \times \hat{z} ~,
\end{eqnarray}
where the term in the square brackets is (minus) the total force
acting on the i-th electron. The time integral of this term over
the duration of the collision
 (be it an electron-impurity or an electron-electron collision) gives the change
  in momentum $\Delta \vec p$ during the collision.
    Thus we see that the change in position is
    $\Delta {\vec r}_{phys}=-2 \alpha \sigma \Delta {\vec p} \times
    \hat{z}$ (this is the so-called ``side-jump"):
 hence the change in energy is correctly given by Eq.~(\ref{epsilon_tylda}).

Kohn and Luttinger\cite{Kohn58}, have shown that the above form of
 the collision integral is correct up to third order in the strength of the
electron-impurity approximation:  this is one order higher than
the Born approximation and should therefore be sufficient to
capture the skewdness of the scattering probability, which arises
from terms beyond the Born approximation.  Notice that the
collision integral does not contain the tempting but ultimately
incorrect  ``Pauli-blocking" factors $1-f_\sigma(\vec k')$.

Similarly, the electron-electron contribution to the
collisional derivative has the form\cite{Amico02}
\begin{eqnarray}\label{coulomb.collision.integral}
\dot f_{\sigma}(\vec k)_{c,e-e}  &\simeq& -\sum_{\vec k' \vec p
\vec p'}W_C(\vec k \sigma,\vec p -\sigma;\vec k' \sigma, \vec p'
-\sigma) \left \{f_{\sigma}(\vec k)f_{-\sigma}(\vec
p)[1-f_{\sigma}(\vec k')] [1-f_{-\sigma}(\vec p' )] \right.
\nonumber\\ &&\left. - f_{\sigma}(\vec k')f_{-\sigma}(\vec
p')[1-f_{\sigma}(\vec k)] [1-f_{-\sigma}(\vec p)] \right\}
\delta_{\vec k +\vec p,\vec k'+\vec p'}
\delta(\widetilde \epsilon_{k\sigma}+\widetilde
\epsilon_{p-\sigma}-\widetilde \epsilon_{k'\sigma} - \widetilde
\epsilon_{p'-\sigma})~,
\end{eqnarray}
where $W_C(\vec k \sigma,\vec p -\sigma;\vec k' \sigma, \vec p'
\-\sigma)$ is the electron-electron scattering rate from $\vec k
\sigma, \vec p -\sigma$  to   $\vec k' \sigma, \vec p' -\sigma$,
and the Pauli factors $f_\sigma(\vec k)$,  $1-f_\sigma(\vec k')$
etc. ensure  that the initial states are occupied and the final
states empty as required by Pauli's exclusion principle. Notice
that, for our purposes, only collisions between electrons of
opposite spins are relevant, since collision between same-spin
electrons conserve the total momentum of each spin component.
Accordingly,  only the former have been retained in
Eq.~(\ref{coulomb.collision.integral}).

\section{Skew-scattering}

Let us, at first, neglect the electron-electron interaction.  From
the general theory  developed, for instance, in
Ref.~\cite{Landau3}, one can easily deduce that  the scattering
amplitude from one impurity in two dimensions  has the form
\begin{equation}\label{scattering.amplitude}
f_{\vec k\vec k',\sigma} = A_{\vec k\vec k'}+\sigma B_{\vec k\vec
k'}(\hat k \times \hat k')_z~,
\end{equation}
where $A_{\vec k\vec k'}$ and $B_{\vec k\vec k'}$ are complex
scattering amplitudes,  and $\hat k$ and $\hat k'$ are the unit vectors in the
directions of $\vec k$ and $\vec k'$ respectively. The second term
on the right-hand side of Eq.~(\ref{scattering.amplitude}) is due
to the spin-orbit interaction.  Squaring the scattering amplitude
and multiplying by the number $N_i$ of independent scattering
centers we arrive at the following expression for the scattering
rate from $\vec k$ to $\vec k'$:
\begin{equation}\label{scattering.probability}
W_{\vec k\vec k',\sigma} = \left[W^s_{\vec k\vec k'}+\sigma
W^a_{\vec k\vec k'} (\hat k \times \hat k')_z\right]
\delta(\epsilon_k-\epsilon_{k'})~,
\end{equation}
where
\begin{equation} \label{def.W+}
W^s_{\vec k\vec k'} =  N_i \left[\vert A_{\vec k\vec k'}\vert^2
+\vert B_{\vec k\vec k'} \vert^2\right]~,
\end{equation}
and
\begin{equation} \label{def.W-}
W^a_{\vec k\vec k'} =  2 N_i Re \left[A_{\vec k\vec k'}B^*_{\vec
k\vec k'}\right]~.
\end{equation}
Here and in the following  $\epsilon_{k} \equiv \frac{\hbar^2
k^2}{2m}$ is the free particle energy. The second term in the
square brackets of Eq.~(\ref{scattering.probability}) depends on
the spin ($\sigma = +1$ or $-1$ for up- or down-spins
respectively) and on the chirality of the scattering (i.e., the
sign of   $(\hat k \times \hat k')_z$).  We will refer to this as
the {\it skew-scattering} term. It should be noted that for a
centrally symmetric scattering potential -- the only case we are
going to consider in this paper --  $W^s_{\vec k\vec k'}$  and
$W^a_{\vec k\vec k'}$ depend only on the magnitude of  the vectors
$\vec k$ and $\vec k'$, which are equal by energy conservation,
and on the angle $\theta$ between them. Furthermore, they are both
symmetric under interchange of $\vec k$ and $\vec k'$ -- the
antisymmetry of the skew-scattering being explicitly brought in by
the factor
    $(\hat k \times \hat k')_z =\sin \theta$.
Thus, in the following, we will often write $W^{s/a}_{\vec k\vec
k'} \equiv W^{s/a}(k,\theta)$, where $W^{s/a}(k,\theta)$ are even
functions of $\theta$.   Notice that the skew-scattering term
vanishes when  the scattering is treated in the second-order Born
approximation~\cite{Landau3}.  Indeed, within this approximation
$A_{\vec k\vec k'}$ is purely real and $B_{\vec k\vec k'}$ is
purely imaginary, so $W^a_{\vec k\vec k'}$ is zero.

The linearized  Boltzmann equation can be solved
exactly  under
the assumption that  $W^s_{\vec k\vec k'}$ and $W^a_{\vec k\vec
k'}$ depend only on the energy $\epsilon_k= \epsilon_{ k'}$ and on
the angle $\theta$ between $\vec k$ and $\vec k'$. The solution
has the form
\begin{equation}\label{Ansatz0}
f_{\sigma}(\vec k) =
f_{0\sigma}(\epsilon_k) - f^\prime_{0\sigma}(\epsilon_k) \hbar \vec k \cdot
\vec V_\sigma (k)~,
\end{equation}
where $\vec V_\sigma (k)$ is proportional to the electric field.  In
view of Eq.~(\ref{epsilon_tylda}) it is convenient to expand

\begin{equation}\label{f0_expand}
f_{0\sigma}({\epsilon}_k) =
f_{0\sigma}(\widetilde{\epsilon}_{k\sigma})-
f'_{0\sigma}(\epsilon_k)(\widetilde{\epsilon}_{k\sigma}-\epsilon_k)~.
\end{equation}
so that our Ansatz~(\ref{Ansatz0}) takes the form
\begin{equation}\label{Ansatz0.1}
f_{\sigma}(\vec k) = f_{0\sigma}(\widetilde \epsilon_{k\sigma})-
2f'_{0\sigma}(\epsilon_k)e\alpha\hbar\sigma
(\vec{E}_{\sigma}\times\hat{z})\cdot \vec{k}-
f^\prime_{0\sigma}(\epsilon_k) \hbar \vec k \cdot \vec V_\sigma
(k)~.
\end{equation}
The advantage of this form is that
the ``zero-order term" $f_{0\sigma}(\widetilde \epsilon_{k\sigma})$
makes no contribution to the collision integral
~(\ref{collision.integral}).   Then, making use of
Eq.~(\ref{scattering.probability}) and discarding terms proportional
to $\alpha W^a$ (which are small since $W^a$ itself is proportional
to $\alpha$)   we arrive at the following form for the linearized
collision integral:
\begin{eqnarray}\label{collision.integral.2}
\dot f_{1\sigma}(\vec k)_{c,imp} &=& -\sum_{\vec k'} W^s_{\vec
k\vec k'} \left\{f_{1\sigma}(\vec k)-f_{1\sigma}(\vec
k')\right\}\delta(\epsilon_k-\epsilon_{k'})\nonumber\\ &-&\sigma
\sum_{\vec k'} W^a_{\vec k\vec k'} (\hat k \times \hat
k')_z\left\{f_{1\sigma}(\vec k)+ f_{1\sigma}(\vec
k')\right\}\delta(\epsilon_k-\epsilon_{k'})\nonumber\\
&+&2\sigma \sum_{\vec k'} W^s_{\vec k\vec k'} f'_{0\sigma}(\epsilon_k)
e\alpha\hbar (\vec{E}_{\sigma}\times\hat{z})\cdot
(\vec{k}-\vec{k'})\delta(\epsilon_k-\epsilon_{k'})~,
\end{eqnarray}
where $f_{1\sigma}(\vec k)\equiv -
f^\prime_{0\sigma}(\epsilon_k) \hbar \vec k \cdot \vec V_\sigma (k)$
is the deviation of the distribution function from unpertubed
equilibrium.

At low temperature, it is not necessary to take into account the
full $k$-dependence of $\vec V_\sigma (k)$, since the derivative
of the Fermi distribution $f^\prime_{0\sigma}(\epsilon_k)$
restricts the range of  $k$ to the vicinity of the
Fermi wave vectors $k_{F\sigma}$. Thus we replace the function
$V_{\sigma}(k)$ by a constant $V_\sigma$, and determine $V_\sigma$
from the consistency condition
\begin{equation}\label{Boltzmann.Equation.2}
-e\sum_{\vec k} \frac{\hbar \vec k}{m}\left[\vec E_\sigma \cdot
\frac{\hbar \vec k}{m}\right]f^\prime_{0\sigma}(\epsilon_k)=
\sum_{\vec k} \frac{\hbar \vec k}{m} \dot f_{1\sigma}(\vec
k)_{c,imp}~.
\end{equation}
Substituting  the collision integral from
Eq.~(\ref{collision.integral.2}) on the right-hand
side of this equation and moving its last term to the left hand side
we arrive at
\begin{eqnarray}\label{Boltzmann.Equation.3}
&&\frac{e n_\sigma}{m}\vec E_\sigma - 2e\sigma\sum_{\vec k k'}\frac{\hbar^2
\vec k}{m}\left[(\vec k-\vec k')W^s_{\vec k\vec k'}\alpha
(\vec{E}_{\sigma}\times\hat{z})
f^\prime_{0\sigma}(\epsilon_k)\delta(\epsilon_k-\epsilon_{k'})\right]
= \sum_{\vec k \vec k'} W^s_{\vec k\vec k'}\frac{\hbar^2 \vec k}{m}
(\vec k - \vec k')\cdot \vec V_\sigma
f^\prime_{0\sigma}(\epsilon_k)
\delta(\epsilon_k-\epsilon_{k'})\nonumber\\&&~~~~~~~~~~~+\sigma
\sum_{\vec k \vec k'}
W^a_{\vec k\vec k'} \frac{\hbar^2 \vec k}{m} (\hat k \times \hat
k')_z  (\vec k + \vec k')\cdot \vec V_\sigma
f^\prime_{0\sigma}(\epsilon_k) \delta(\epsilon_k-\epsilon_{k'})~,
\end{eqnarray}
where $n_\sigma = \frac{k_{F\sigma}^2}{4\pi}$ is the density of
$\sigma$-spin carriers and $k_{F\sigma}$ is the corresponding
Fermi wave vector. The first term on the right-hand side of
Eq.~(\ref{Boltzmann.Equation.3}) is parallel to $\vec V_\sigma$,
while the second term is orthogonal to it. Then a simple
calculation leads to the following expression for  $\vec V_\sigma$
in terms of the electric field:
\begin{equation}\label{resistivity.matrix1}
- \frac{e}{m} \vec E_{\sigma}- \frac{2e\alpha\sigma
(\vec{E}_{\sigma}\times\hat{z})}{\tau_{\sigma}} = \frac{\vec
V_\sigma}{\tau_\sigma} + \sigma \frac{\vec V_\sigma \times\hat
z}{\tau^{ss}_\sigma}~,
\end{equation}
and its inverse to first order in $\alpha$ is
\begin{equation}\label{Vsigma}
\vec V_\sigma = \frac{-e\tau_\sigma}{m}\left[\vec E_\sigma -
\sigma \frac{\tau_\sigma}{\tau_\sigma^{ss}}
   \vec E_\sigma \times \hat z \right]-2e\alpha \sigma
(\vec{E_{\sigma}}\times \hat{z})~,
\end{equation}
where
\begin{equation}\label{tau.regular}
\frac{1}{\tau_\sigma} = -\frac{m {\cal A}}{4 \pi^2 \hbar^2
\epsilon_{F\sigma}}\int_0^\infty d\epsilon  \epsilon
f^\prime_{0\sigma}(\epsilon)\int_0^{2\pi} d \theta~ W^s(\epsilon,
\theta) (1-\cos \theta)~,
\end{equation}
and
\begin{equation}\label{tau.anomalous}
\frac{1}{\tau^{ss}_\sigma} = -\frac{m {\cal A}}{4 \pi^2 \hbar^2
\epsilon_{F\sigma}}\int_0^\infty d\epsilon \epsilon
f^{\prime}_{0\sigma}(\epsilon)\int_0^{2\pi} d \theta~
W^a(\epsilon,\theta)\sin^2 \theta~.
\end{equation}
In the above equations $\epsilon_{F\sigma}=
\frac{k_{F\sigma}^2\hbar^2}{2m}$ is the Fermi energy for spin
$\sigma$. In the limit of zero temperature the derivative of the
Fermi function reduces to $f^{\prime}_{0\sigma}(\epsilon) \simeq -
\delta (\epsilon-\epsilon_{F\sigma})$ and the above formulas
simplify as follows:
\begin{equation}\label{tau.regular.2}
\frac{1}{\tau_\sigma} \stackrel{T \to 0}{\simeq} \frac{m {\cal
A}}{4 \pi^2 \hbar^2}\int_0^{2\pi} d \theta~ W^s(k_F, \theta)
(1-\cos \theta)~,
\end{equation}
and
\begin{equation}\label{tau.anomalous.2}
\frac{1}{\tau^{ss}_\sigma} \stackrel{T \to 0}{\simeq}  \frac{m
{\cal A}}{4 \pi^2 \hbar^2}\int_0^{2\pi} d \theta~ W^a(k_F, \theta)
\sin^2 \theta~
\end{equation}

\begin{figure}[thb]
\vskip 0.27 in
\includegraphics[width=3.4in]{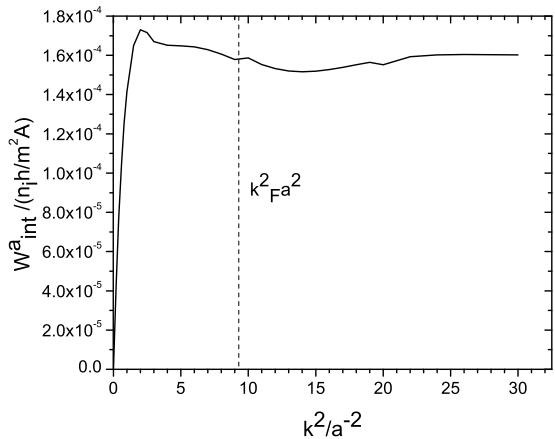}
\caption{Antisymmetric scattering rate $W^{a}_{int}=\int_0^{2\pi}
d \theta~ W^a(\epsilon,\theta)\sin^2 \theta~$ in units of $n_i
h/m^2\cal A$ (see Eq.~(\ref{scattering.probability.minus3})) as a
function of $k^2$ for a model circular well attractive potential
$V_0=-5$meV and radius $a=9.45$nm (described in the Appendix). We
choose the parameters typical for the experimental 2DEG confined
in Al$_{0.1}$Ga$_{0.9}$As quantum well i.e. % $n=10^{12}$cm$^{-2}$,\\
density of electrons and impurities $n_{2D}=n_i=2.0\times
10^{12}$cm$^{-2}$, $m=0.074$m$_e$, and mobility $\bar{\mu}
=0.1$m$^2/$Vs. The effective spin-orbit coupling  $\alpha\hbar=
4.4{\AA}^2$ in accordance with \cite{Winkler2003}.}
\end{figure}
Fig.1 shows the antisymmetric scattering rate $\int_0^{2\pi} d
\theta~ W^a(\epsilon,\theta)\sin^2 \theta~$ calculated numerically
using Eq.~(\ref{scattering.probability.minus3}) for a model
impurity potential Eq.~(\ref{scatter_V}) presented in Appendix,
and for the typical experimental parameters \cite{Sih05}.

\section {Spin-Hall current and side-jump}
The quantity $\vec V_\sigma$ obtained in Eq.~(\ref{Vsigma}),
determines the non-equilibrium distribution,  according to
Eq.~(\ref{Ansatz0}).  We now use this distribution to calculate
the current density.  In order to do this correctly, however, we
must remember that the spin-orbit interaction alters the relation
between the velocity and the canonical momentum. The correct expression for the velocity is given in  Eq.~(\ref{v_phys}) and in the absence of electron-electron interactions takes the form:
\begin{equation}
\vec v_i  = \frac{\vec p_i}{m}+ 2\alpha[\vec \nabla V_{ei}(\vec r_i)+e \vec E] \times \vec \sigma_i~.
\end{equation}
The second term on the right hand side of this equation contains
the net force resulting from the combined action of the external
electric field and the impurity potential on the $i$-th electron.
This force must vanish when averaged in a non-equilibrium
steady-state ensemble, since the average value of the momentum
must be stationary. Also, the same is true in the presence of
electron-electron interactions, provided the Coulomb force is
included. Thus, we arrive at the simple result
\begin{equation}\label{jtotal}
\vec j_{\sigma}= -e n_\sigma \vec V_\sigma~.
\end{equation}
Combining this with Eq.~(\ref{Vsigma}) we obtain
the complete  relation between electric field and current density:
\begin{equation}\label{jno_ee}
\vec j_\sigma= \frac{n_\sigma e^2 \tau_\sigma}{m}
   \left[\vec E_\sigma - \sigma \frac{\tau_\sigma}{\tau_\sigma^{ss}}
   \vec E_\sigma \times \hat z \right]
   +2e^2\alpha \sigma n_{\sigma} \vec
E_\sigma \times \hat z~.
\end{equation}
The last term on the right-hand side of this expression is known
in the literature as the ``side-jump" contribution to the current
density \cite{Nozieres}.  It comes from the use of $\widetilde \epsilon_k$ rather than $\epsilon_k$ in the $\delta$-function of conservation of energy  -- see discussion in the paragraph following Eq.~(\ref{epsilon_tylda}). Inverting Eq.~(\ref{jno_ee}) we obtain the formula for the
electric field in terms of the current densities:
\begin{equation}\label{electric_so}
\vec E_\sigma=\rho^D_{\sigma}\vec
j_{\sigma}+\sigma[\rho^{ss}_{\sigma}-
\lambda_{\sigma}\rho^D_{\sigma}]\vec j_{\sigma}\times \hat{z}
\end{equation}
where $\rho^D_{\sigma} =\frac{m}{n_{\sigma}e^2\tau_{\sigma}}$ is
the Drude resistivity, $\rho^{ss}_{\sigma}
=\frac{m}{n_{\sigma}e^2\tau^{ss}_{\sigma}}$, $\lambda_{\sigma}
=\frac{2m\alpha}{\tau_{\sigma}}$. Hence the resistivity tensor,
written in the basis $x_{\uparrow}$, $y_{\uparrow}$,
$x_{\downarrow}$, $y_{\downarrow}$, has the following form:
\begin{equation}\rho \label{diagonal.resistivity}=
\left(%
\begin{array}{cccc}
     \rho^D_{\uparrow}
   & \rho^{ss}_{\uparrow}-
\lambda_{\uparrow}\rho^D_{\uparrow} & 0 & 0
\\-\rho^{ss}_{\uparrow}+
\lambda_{\uparrow}\rho^D_{\uparrow}& \rho^D_{\uparrow} & 0 & 0 \\
     0 & 0 & \rho^D_{\downarrow}
& -\rho^{ss}_{\downarrow}+ \lambda_{\downarrow}\rho^D_{\downarrow}
\\
     0 & 0 & \rho^{ss}_{\downarrow}-
\lambda_{\downarrow}\rho^D_{\downarrow}
& \rho^D_{\downarrow}\\
\end{array}%
\right)
\end{equation}
The diagonal part of the resistivity reduces to the Drude formula
$\rho_{\sigma}^D =\frac{m}{ne^2\tau_{\sigma}}$ as expected.  The
spin-orbit interaction is entirely responsible for the appearance
of an off-diagonal (transverse) resistivity.   The latter consists of two
competing terms associated with side-jump
($\lambda_{\sigma}\rho^D_{\sigma}$) and skew-scattering
($\rho^{ss}_{\sigma}$), as seen in
Eq.~(\ref{diagonal.resistivity}). Hence our expression for the
transverse resistivity is different from the expression presented
in previous papers (see for example Ref.\cite{Zhang00}) where only
the side-jump contribution  appears in the final formulas.  Notice
that, at this level of approximation, the resistivity is diagonal
in the spin indices.

\section{Spin Coulomb drag}
Up to this point we have ignored electron-electron scattering
processes, as well as scattering processes that might flip the
spin of the electrons. As discussed in Ref.~(\cite{Amico02}) these
processes are important because they couple the up- and down-spin
components of the current density, giving rise to off-diagonal
elements, $\rho^{\alpha\beta}_{\uparrow\downarrow}$, of the
resistivity tensor. The Coulomb interaction, in particular, leads
to the phenomenon of the spin Coulomb drag (SCD) -- a form of
friction caused by the relative drift motion of spin up and spin
down electrons, and the consequent transfer of momentum between
them \cite{Amico02,Amico03,Flensberg01,Orenstein05}.
  Both Coulomb and
spin-flip scattering can be included in our formulation as
additional contributions to the collisional derivative $\dot
f_{1\sigma}(\vec k)_{c}$ (see Ref.~(\cite{Amico02})). In
particular, the Coulomb contribution is given by
Eq.~(\ref{coulomb.collision.integral}). Substituting
Eq.~(\ref{Ansatz0.1}) into Eq.~(\ref{coulomb.collision.integral})
and including the first order corrections from electron-electron
interactions to the spin-orbit Hamiltonian, we arrive at the
following expression for the Coulomb collision integral:
\begin{eqnarray}\label{coulomb.collision.integral3}
\dot f_{\sigma}(\vec k)_{c,e-e}  \simeq -\frac{1}{k_BT}\sum_{\vec k' \vec p
\vec p'}W_C(\vec k \sigma,\vec p -\sigma;\vec k' \sigma, \vec p'
-\sigma) [\hbar \vec V_{\sigma}-\hbar \vec V_{-\sigma}+ 2e
\alpha\hbar
 \sigma (E_\sigma+E_{-\sigma}) \times \hat{z}](\vec k-\vec k')\\ \nonumber
 f_{0\sigma}(\epsilon_ k)f_{0-\sigma}(\epsilon_ p)f_{0
\sigma}(-\epsilon_{k'})f_{0-\sigma}(-\epsilon_{p'})
 \delta_{\vec k +\vec p,\vec k'+\vec p'} \delta(\epsilon_{k\sigma}+\epsilon_{p-\sigma}-\epsilon_{k'\sigma} - \epsilon_{p'-\sigma})~,
\end{eqnarray}
where $T$ is the temperature, $k_B$ is the Boltzmann constant, and we have made use of the identity $f_{0\sigma}(\epsilon_
k)f_{0-\sigma}(\epsilon_ p)[1-f_{0
\sigma}(\epsilon_{k'})][1-f_{0-\sigma}(\epsilon_{p'})]=[1-f_{0\sigma}(\epsilon_
k)][1-f_{0-\sigma}(\epsilon_ p)]f_{0
\sigma}(\epsilon_{k'})f_{0-\sigma}(\epsilon_{p'})$  for
$\epsilon_{k\sigma}+ \epsilon_{p-\sigma}- \epsilon_{k'\sigma} -
\epsilon_{p'-\sigma}=0$.

Eq.~(\ref{coulomb.collision.integral3})  is now inserted into the ``consistency condition"~(\ref{Boltzmann.Equation.2}), and the resulting sum over momenta is expressed in terms of the spin drag coefficient $\gamma$, i.e. the rate of momentum transfer between up- and down-spin electrons, according to the formula\cite{Amico02}
\begin{eqnarray} \label{gamma.Boltzmann}
\gamma &=& \frac{n}{n_\sigma n_{-\sigma}}\sum_{\vec k \vec k' \vec p
\vec p'}W_C(\vec k \sigma,\vec p -\sigma;\vec k' \sigma, \vec p'
-\sigma) \frac {(\vec k - \vec k')^2}{4 mk_BT}
f_{0\sigma}(\epsilon_ k)f_{0-\sigma}(\epsilon_ p)f_{0
\sigma}(-\epsilon_{k'})f_{0-\sigma}(-\epsilon_{p'})
\nonumber\\&\times& \delta_{\vec k +\vec p,\vec k'+\vec p'}  \delta(\epsilon_{k\sigma}+\epsilon_{p-\sigma}-\epsilon_{k'\sigma} - \epsilon_{p'-\sigma}).
\end{eqnarray}

Then equation~(\ref{resistivity.matrix1}) for $\vec V_{\sigma}$ is
modified  as follows:
\begin{equation}\label{SD1}
- \frac{e}{m} \vec E_{\sigma}-
2\frac{e\alpha\sigma(\vec{E}_{\sigma}\times
\hat{z})}{\tau_{\sigma}}= \frac{\vec V_\sigma}{\tau_\sigma} +
\sigma \frac{ \vec V_\sigma \times \hat z
}{\tau^{ss}_\sigma}-\frac{n_{-{\sigma}}\gamma}{n_{\sigma}ne}\vec{j}_{\sigma}
+\frac{\gamma}{ne}\vec{j}_{-{\sigma}}+2\gamma\frac{n_{-\sigma}}{n}e\alpha\sigma
(\vec{E}_{\sigma}+\vec{E}_{-\sigma})\times \hat{z}
+\frac{1}{\tau'_{\sigma}n_{\sigma}e}\vec{j}_{-{\sigma}},
\end{equation}
The third and fourth terms on the right hand side (r.h.s.) of this
equation are connected with the SCD and $\gamma$ is the spin-drag
coefficient calculated in Ref.~\cite{Amico03}. The fifth term on
r.h.s. of Eq.~\ref{SD1} comes from the side-jump effect in Coulomb scattering.   The last term in Eq.~(\ref{SD1}) is associated with
spin-flip collision processes, characterized by the spin
relaxation time $\tau^\prime_\sigma$.  We include it for completeness, even though spin-flip effects are expected to be  small in n-doped semiconductors. The  current  is now given by the full velocity operator of Eq.~(\ref{v_phys}), but thanks to the force balance condition in the steady state it reduces again to
the simple form
 \begin{equation}\label{jSD_sigma2} \vec j_{\sigma}= -e n_\sigma
\vec V_\sigma ~.
\end{equation}
Combining (Eq.~\ref{SD1}) with
(Eq.~\ref{jSD_sigma2}) leads to the  resistivity tensor,
which in the basis of $x_{\uparrow}$, $y_{\uparrow}$,
$x_{\downarrow}$, $y_{\downarrow}$ has the form:
\begin{equation} \label{Coulomb.Rho.Matrix}\rho= \left(
\begin{array}{cccc}
\rho^D_{\uparrow} + \rho^{SD}n_{\downarrow}/n_{\uparrow} &
\rho^{ss}_{\uparrow}-
\lambda_{\uparrow}\rho^D_{\uparrow}+A^{\gamma\alpha}_{\uparrow} &
-\rho^{SD}-\rho^\prime_{\uparrow}& B^{\gamma\alpha}_{\uparrow}
\\-\rho^{ss}_{\uparrow}+
\lambda_{\uparrow}\rho^D_{\uparrow}-A^{\gamma\alpha}_{\uparrow}&
\rho^D_{\uparrow} +\rho^{SD}n_{\downarrow}/n_{\uparrow} &
-B^{\gamma\alpha}_{\uparrow} & -\rho^{SD}-\rho^\prime_{\uparrow}
\\ -\rho^{SD}-\rho^\prime_{\downarrow} & -B^{\gamma\alpha}_{\downarrow} & \rho^D_{\downarrow} +
\rho^{SD}n_{\uparrow}/n_{\downarrow}& -\rho^{ss}_{\downarrow}+
\lambda_{\downarrow}\rho^D_{\downarrow}-A^{\gamma\alpha}_{\downarrow}\\
 B^{\gamma\alpha}_{\downarrow}  & -\rho^{SD}-\rho^\prime_{\downarrow}  &
 \rho^{ss}_{\downarrow}-\lambda_{\downarrow}\rho^D_{\downarrow}+A^{\gamma\alpha}_{\downarrow}
&\rho^D_{\downarrow}+ \rho^{SD}n_{\uparrow}/n_{\downarrow}\\
\end{array}%
\right)
\end{equation}
where $\rho^{SD} =m\gamma/ne^2$ is the spin Coulomb drag
resistivity and
$\rho^\prime_{\sigma}=m/n_{\sigma}e^2\tau^\prime_{\sigma}$
(recall that $\lambda_\sigma = \frac{2 m \alpha}{\tau_\sigma}$ is a dimensionless quantity).
$A^{\gamma\alpha}_{\sigma}$ and $B^{\gamma\alpha}_{\sigma}$
represent the terms of the first order in electron-electron
coupling $\gamma$ and in SO coupling $\alpha$ and are defined as
follows:
$A^{\gamma\alpha}_{\sigma}=-\lambda_{\sigma}\rho_{SD}n_{-\sigma}/n_{\sigma}
+2m^*\alpha\gamma[-n_{-\sigma}\rho^D_{\sigma}/n
+(n_{-\sigma}/n-n^2_{-\sigma}/nn_{\sigma})\rho^{SD}]$ and
$B^{\gamma\alpha}_{\sigma}=\lambda_{\sigma}\rho_{SD}+2m^*\alpha\gamma[-n_{-\sigma}\rho^D_{-\sigma}/n
+(n_{-\sigma}/n-n_{\sigma}/n)\rho^{SD}]$.
 Notice that the resistivity satisfies the following symmetry relations:
\begin{equation}\label{symmetry1}
\rho^{\beta \beta'}_{\sigma \sigma}= -\rho^{\beta' \beta}_{\sigma
\sigma}
\end{equation}
\begin{equation}\label{symmetry2}
\rho^{\beta \beta'}_{\sigma -{\sigma}}= \rho^{\beta'
\beta}_{-{\sigma} \sigma}
\end{equation}
where upper indices $\beta$ and $\beta'$ denote directions, and
the lower ones spin orientations. From
Eq.~(\ref{Coulomb.Rho.Matrix}) one clearly sees that the spin
Coulomb drag and spin-flip processes couple the spin components of
the longitudinal resistivity. Further, the spin-Coulomb drag
corrections to the spin-orbit hamiltonian ($\alpha\gamma$
corrections) result in the $\rho^{\beta\beta'}_{\sigma -{\sigma}}$
terms, i.e. in the transverse resistivity which couples opposite
spins. Also, the SCD renormalizes the longitudinal spin-diagonal
components $\rho^{\beta \beta}_{\sigma -\sigma}$ of the
resistivity, in such a way as to satisfy Galilean invariance.

The spin-flip collision processes come as a phenomenological term
which could have origin for example in some random magnetic field,
which does not appear in the original spin-conserving Hamiltonian
(Eq.~(\ref{H})). The spin-flip relaxation time,
$\tau^\prime_\sigma$ is given by the microscopic
expression~\cite{Amico02}
\begin{equation} \label{tauprime}
\frac{1}{\tau^\prime_\sigma} = - \frac{m {\cal A}}{8 \pi^2 \hbar^2
\epsilon_{F-{\sigma}}}  \int_{0}^{2\pi}d \theta W^{sf}_{\sigma,
-{\sigma}} (\epsilon_k,\theta) \cos \theta ~,
\end{equation}
where $W^{sf}_{\sigma, -{\sigma}}$  denotes the spin-flip
scattering rate  from spin $\sigma$ to the opposite spin
orientation $-{\sigma}$.\footnote{Strictly speaking spin-flip
scattering also renormalizes the ordinary momentum relaxation time
$\tau_\sigma$. This small correction is neglected here: it can
easily be taken into account if needed.}. Since the relaxation
time for spin-flip processes $\tau^\prime$ is very long,
\cite{Amico02} the SCD normally controls the coupling between the
spin components, except at the very lowest temperatures (the spin
drag rate $\gamma$ vanishes as $T^2\ln T$ while the spin-flip rate
remains constant).  Spin-flip processes  will therefore be omitted
henceforth.
\begin{figure}[thb]
\vskip 0.27 in
\includegraphics[width=3.6in]{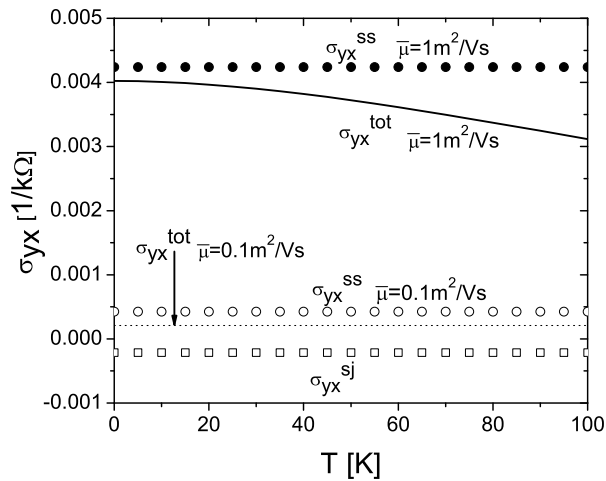}
\caption{Spin Hall conductivity as a function of temperature.
$\sigma_{yx}^{sj}$(open squares), $\sigma_{yx}^{ss}$ (open and
close circles) are the side-jump and skew-scattering
contributions, respectively, to the conductivity in the absence of
electron-electron interactions, $\sigma_{yx}^{tot}$ is the total
spin conductivity when electron-electron interactions are taken
into account. We choose the parameters typical for the
experimental 2DEG confined in Al$_{0.1}$Ga$_{0.9}$As quantum well
i.e. density of electrons and impurities $n_{2D}=n_i=2.0\times
10^{12}$cm$^{-2}$, $m=0.074$m$_e$, and two sets of mobilities and
relaxations times: $\bar{\mu} =0.1$m$^2/$Vs,  $\tau =4\times
10^{-5}$ns $\tau_{ss} = 0.02$ns and $\bar{\mu} =1$m$^2/$Vs $\tau
=4\times 10^{-4}$ns $\tau_{ss} = 0.2$ns. The effective spin-orbit
coupling  $\alpha\hbar= 4.4{\AA}^2$ in accordance with
\cite{Winkler2003}. We used the model potential (see appendix)
where an effective impurity radius $a=9.45$nm, the height of
attractive impurity potential $V_0=-5$meV for $\bar{\mu}
=0.1$m$^2/$Vs and $V_0=-1.6$meV for $\bar{\mu} =1$m$^2/$Vs.}
\end{figure}

In a paramagnetic material there is symmetry between up and down
spin densities, mobilities, etc. and one  can easily separate the
spin and charge degrees of freedom. Then combining the
Eq.~(\ref{SD1}) and  Eq.~(\ref{jSD_sigma2}) simplifies to:
\begin{equation}\label{Ec}
\vec E_c = \rho^{D}\vec j_c+2(\rho^{ss}-\lambda
\rho^D-\lambda\rho_{SD})\vec j_s\times\hat{z}~,
\end{equation}
\begin{equation}\label{Es}
\vec E_s=4(\rho^{SD}+\rho^{D})\vec j_s+2(\rho^{ss}-\lambda
\rho^D-\lambda\rho_{SD})\vec j_c\times\hat{z}~,
\end{equation}
where we omit spin-flip processes, and the charge/spin components
of the electric field are defined as  $\vec{E}_c
=\frac{\vec{E}_{\uparrow}+\vec{E}_{\downarrow}}{2}$, $\vec{E}_s
=\vec{E}_{\uparrow}-\vec{E}_{\downarrow}$, and the charge and spin
currents are $\vec j_c =\vec j_{\uparrow}+\vec j_{\downarrow}$ and
$\vec j_{s} = \frac{\vec j_{\uparrow}-\vec j_{\downarrow}}{2}$,
respectively. The spin-Coulomb drag renormalizes the longitudinal
resistivity only in the spin channel. This is a consequence of the
fact that the net force exerted by spin-up electrons on spin-down
electrons is proportional to the difference of their drift
velocities, i.e. to the spin current. Additionally, the
electron-electron corrections  to the spin-orbit interactions
renormalize the transverse resistivity in the charge and spin
channels, so the Onsager relations between spin and charge
channels hold.  Under the assumption that the electric field is in the $x$
direction has the same value for spin up and spin down we see that
Eqs.~(\ref{Ec}) and~(\ref{Es}) yield the following formula for
the spin current in $y$ direction:
\begin{equation}\label{Jsy1}
{j^{s}}^{z}_y =
\left[\frac{\rho^{ss}/2(\rho^D)^2}{1+\rho^{SD}/\rho^D}
-\frac{\lambda}{2\rho_{D}}\right] E_x
\end{equation}
The first term in the square brackets  is associated
with the skew-scattering, while the second is the side-jump
contribution. Notice that the side-jump conductivity
$-\frac{\lambda}{2\rho_{D}} = - \alpha n e^2$  depends
neither on the strength of disorder nor on the strength of the
electron-electron interaction. In contrast, the spin-Coulomb drag
{\it reduces} by factor $1+\rho^{SD}/\rho^D$ the skew-scattering
term, so the total  spin Hall conductivity and the spin Hall
current are reduced as well.

The temperature dependence of  the spin conductivities
$\sigma_{yx}={j^s}^z_y/E_x$ for two different mobilities and for
the parameters of the recent experiment  on the 2DEG in
Al$_{0.1}$Ga$_{0.9}$As \cite{Sih05} is presented in Fig.2. We used
Eqs.~(\ref{scattering.probability.plus3}) and
(\ref{scattering.probability.minus3}) in the appendix to calculate
the scattering rates. The procedure is as follows. First we find
the scattering rate $\tau_{\sigma} (=\tau_{-\sigma})$ from the
mobility. Then using Eqs.~(\ref{scattering.probability.plus3}) and
(\ref{tau.regular}) we estimate the strength of impurity potential
$V_0$. Finally, we compute the skew-scattering rate by numerically
evaluating Eq.~(\ref{scattering.probability.minus3}). In general
the skew-scattering rate of Eq.~(\ref{tau.regular}) is temperature
dependent through the Fermi distribution. However, in a wide range
of mobilities  the impurity potential $V_0$ is much smaller than
the Fermi energy and the scattering rate does not change with
energy around $E_F$, i.e. $[\partial
W^{a}(\epsilon,\Theta)/\partial \epsilon]_{\epsilon=E_F}\approx 0$
(see Fig.1). Since the temperature dependence of  the
skew-scattering rate comes from the energy dependence of
$W^{a}(\epsilon,\Theta)$,    we see that the temperature
dependence of skew-scattering term is very weak. In Fig.2,
$\sigma_{yx}^{sj}$ is the side-jump contribution to the
conductivity in the absence of electron-electron interactions
(found from the last term of Eq.~\ref{jno_ee}). $\sigma_{yx}^{ss}$
is the skew-scattering conductivity in the absence of
electron-electron interactions (found from second term of
Eq.~(\ref{jno_ee})). $\sigma_{yx}^{tot}$ is the total off-diagonal
conductivity modified by the spin Coulomb drag term, as given by
Eq.~(\ref{Jsy1}). For $T=0$  the total spin conductivity is the
sum of two contributions: skew-scattering proportional to
$\rho_{ss}/2\rho_D^2$ and side-jump proportional to
$\lambda/2\rho_D = -\alpha ne^2$. For an attractive electron-impurity potential
($V_0<0$) the two contributions have opposite signs,  consistent
with previous theoretical estimates in
Refs.~\cite{Engel05,Sarma05}. The ratio of the skew-scattering to
side-jump terms depends on the mobility. The skew-scattering
conductivity scales as $\mu$ while the side-jump is independent of
mobility. As a consequence the skew-scattering conductivity
dominates for high mobilities while the side-jump dominates at low
mobilities. The different ratios of skew-scattering to side-jump
conductivities reported in recent theoretical papers
\cite{Engel05,Sarma05} result from choosing different mobilities
(see also Fig.2). The spin-Coulomb drag is the only temperature
dependent contribution in this calculations: it scales as $T^2\ln
T$ for $T<T_F$ and reduces the absolute value of spin conductivity
and spin current. Moreover, the reduction of spin-Hall effect by
spin-Coulomb drag depends strongly on the Drude resistivity. Hence
the reduction of spin-Hall conductivity will be of the order of a
few percent for low mobilities (invisible change in $\sigma_{yx}$
for $\bar{\mu}=0.1$m$^2$/Vs in Fig.2) and of the order of
25\%-50\% for high mobilities (see $\sigma_{yx}$ for
$\bar{\mu}=1$m$^2$/Vs in Fig.2).

Thus far, as stated in the introduction, we have only considered
the extrinsic spin Hall effect.  What about the intrinsic effect?
In recent experiments by Sih et al. \cite{Sih05} on a 2DEG
confined in an (110) AlGaAs quantum well, three different
contributions to the spin Hall conductivity are present in
principle: the impurities, the linear-in-$\vec k$ SO Rashba field,
and the cubic-in-$\vec k$ SO Dresselhaus field which is
perpendicular to the Rashba field. Since in (110) quantum wells,
the Dresselhaus field is in the [110] direction; we do not expect
any spin-Hall current or spin-Hall accumulation connected with
this term in the plane of the quantum well. Further, it has been
established that the spin-Hall conductivity in an infinite 2DEG
with {\it only} linear spin-orbit interactions of Rashba or
Dresselhaus type vanishes for arbitrarily weak disorder
\cite{Halperin04,Khaetskii04,Schwab04,Hankiewicz_unpublished}.
Hence we believe that the extrinsic contribution $\sigma_{yx}\sim
4\times 10^{-4}1/k\Omega$ is the dominant one in the referenced
paper. Also, the theoretical estimates of extrinsic
~\cite{Engel05} and intrinsic ~\cite{Bernevig04} contributions in
the experiments on three-dimensional n-doped GaAs ~\cite{Kato04}
show that the extrinsic contribution is an order of magnitude
larger than the intrinsic one.

\section {Spin accumulations}
Let us now study the influence of spin-Coulomb drag on the spin
accumulation. This is relevant to the interpretation of the
experiments by Sih {\it et al.} \cite{Sih05} and Kato {\it et
al.}\cite{Kato04}. We consider a very long conductor in the form
of a bar of length $L$ in the $x$ direction and narrow width $W$
in the $y$ direction. A charge current flows only in the $x$
direction. The $y$-components of the current $j_\sigma^y$, with
$\sigma = \uparrow$ and $\downarrow$ add up to zero everywhere and
individually vanish on the edges of the system, i.e. $
j_\sigma^y=0$ at $y = \pm W/2$. In order to satisfy the boundary
conditions the system cannot remain homogeneous in the
$y$-direction.  A position-dependent spin density, known as {\it
spin accumulation}  develops across the bar, and is reflected in
non-uniform chemical potentials $\mu_\sigma(y)$. In the steady
state regime the spatial derivative of the spin-current in the
$y$-direction must exactly balance the relaxation of the spin
density due to spin-flip processes.  This condition leads to the
following equation for the spin chemical potentials \cite{Fert93}
\begin{equation}\label{spin_diffusion}
\frac{d^2[\mu_{\sigma}(y)-\mu_{-\sigma}(y)]}{d^2y}
=\frac{\mu_{\sigma}(y)-\mu_{-{\sigma}}(y)}{L_s^2}~,
\end{equation}
where $L_s$ is the spin diffusion length. The solution of this
equation is:
\begin{equation}\label{spin_diffusion1a}
\mu_{\sigma}(y)-\mu_{-{\sigma}}(y)=Ce^{\frac{y}{L_s}}+
C'e^{-\frac{y}{L_s}}~,
\end{equation}
where C, C' are constants to be determined by the boundary
conditions $j^y_{\pm \sigma}(\pm W/2) =0$. The effective electric
field in the $y$- direction is $eE^y_{\sigma}=d\mu_{\sigma}/dy$.
Thus,
\begin{equation}\label{electric_field2}
-e(E^y_{\uparrow}-E^y_{\downarrow})
=\frac{d(\mu_{\downarrow}-\mu_{\uparrow})}{dy}=
\frac{C'}{L_s}e^{-y/L_s}-\frac{C}{L_s}e^{y/L_s}~.
\end{equation}
Using the boundary conditions at $y=\pm W/2$ we obtain
\begin{equation}\label{Eyup1}
E^y_{\uparrow}(\pm
W/2)=\rho^{yx}_{\uparrow\uparrow}j^x_{\uparrow}+\rho^{yx}_{\uparrow\downarrow}j^x_{\downarrow}
\end{equation}
\begin{equation}\label{Eydown1}
E^y_{\downarrow}(\pm
W/2)=\rho^{yx}_{\downarrow\uparrow}j^x_{\uparrow}+\rho^{yx}_{\downarrow\downarrow}j^x_{\downarrow}
\end{equation}
Making use of  Eqs.~(\ref{electric_field2}-\ref{Eydown1})  to
eliminate the electric field, we obtain  the following set of
equations:
\begin{equation}\label{coefficient1}
\frac{C'}{L_s}e^{-W/2L_s}-\frac{C}{L_s}e^{W/2L_s}=\frac{C'}{L_s}e^{W/2L_s}-\frac{C}{L_s}e^{-W/2L_s}=
-e[\rho^{yx}_{\uparrow\uparrow}j^x_{\uparrow}+\rho^{yx}_{\uparrow\downarrow}j^x_{\downarrow}-
\rho^{yx}_{\downarrow\uparrow}j^x_{\uparrow}-\rho^{yx}_{\downarrow\downarrow}j^x_{\downarrow}]~,
\end{equation}
which gives immediately the solution $C= -C'$  and
\begin{equation}\label{coefficient3}
C'=\frac{-eL_s[\rho^{yx}_{\uparrow\uparrow}j^x_{\uparrow}+\rho^{yx}_{\uparrow\downarrow}j^x_{\downarrow}-
\rho^{yx}_{\downarrow\uparrow}j^x_{\uparrow}-\rho^{yx}_{\downarrow\downarrow}j^x_{\downarrow}]}{2\cosh(W/2L_s)}~.
\end{equation}
Thus, the formula for the spin accumulation is
\begin{equation}\label{spin_diffusion1b}
\mu_{\uparrow}(y)-\mu_{\downarrow}(y)=
\frac{eL_s[\rho^{yx}_{\uparrow\uparrow}j^x_{\uparrow}+\rho^{yx}_{\uparrow\downarrow}j^x_{\downarrow}-
\rho^{yx}_{\downarrow\uparrow}j^x_{\uparrow}-\rho^{yx}_{\downarrow\downarrow}j^x_{\downarrow}]\sinh(y/L_s)}{
\cosh(W/2L_s)}~.
\end{equation}
Finally, upon substituting the matrix elements of the resistivity
from Eq.~(\ref{Coulomb.Rho.Matrix}) we find
\begin{equation}\label{spin_diffusion2}
\mu_{\uparrow}(y)-\mu_{\downarrow}(y)= \frac{-eL_s \left[
j^{x}_{\uparrow}(\rho^{ss}_{\uparrow}-\lambda_{\uparrow}\rho^D_{\uparrow}+A^{\gamma\alpha}_{\uparrow}+
B^{\gamma\alpha}_{\uparrow})
+j^{x}_{\downarrow}(\rho^{ss}_{\downarrow}
-\lambda_{\downarrow}\rho^D_{\downarrow}+A^{\gamma\alpha}_{\downarrow}+
B^{\gamma\alpha}_{\downarrow})\right]
\sinh(y/L_s)}{\cosh(W/2L_s)}~.
\end{equation}

The spin-Coulomb drag modifies the spin accumulation in three
different ways: (i) the spin Coulomb drag resistivity appears
directly in the terms $A^{\gamma\alpha}$ and $B^{\gamma\alpha}$,
defined after Eq.~(\ref{Coulomb.Rho.Matrix});  (ii) the values of
the spin components of the longitudinal current $j_{\sigma}^x$
are, in general, affected by the SCD; (iii) the spin diffusion
length $L_s$ is significantly reduced by the spin Coulomb
drag\cite{Orenstein05} as seen from the formula \cite{Amico02}
\begin{equation}\label{Ls}
L_s=\frac{\chi_0}{\chi_s}\frac{L_c}{1+\rho_{SD}/\rho_D}~,
\end{equation}
where $\chi_0$ is the susceptibility, $\chi_S$ is the spin
susceptibility, and $L_c$ is the density diffusion length.

In paramagnetic materials Eq.~(\ref{spin_diffusion2}) simplifies to:
\begin{equation}\label{spin_diffusion4}
\mu_{\uparrow}(y)-\mu_{\downarrow}(y)=\frac{-2eL_sj_x[\rho^{ss}-
\lambda\rho^D-\lambda\rho_{SD}]\sinh(y/L_s)}{\cosh(W/2L_s)}=
\frac{-2eL_sE_x[\rho^{ss}-
\lambda\rho^D-\lambda\rho_{SD}]\sinh(y/L_s)}{\rho_D\cosh(W/2L_s)}~.
\end{equation}
where $j_x =E_x/\rho_D$ indeed is independent of $\rho_{SD}$.
Further the Eq.~(\ref{spin_diffusion4}) at the edges of sample for
$L=W/2$ gives:
\begin{equation}\label{spin_diffusion5}
\mu_{\uparrow}(W/2L_s)-\mu_{\downarrow}(W/2L_s)=
-2eL_sj_x[\rho^{ss}-
\lambda\rho^D-\lambda\rho_{SD}]\tanh(W/2L_s)~.
\end{equation}
The three terms in the square bracket of
Eq.~(\ref{spin_diffusion5}) are the skew-scattering term, the
side-jump contribution and the electron-electron correction. The
last term reduces the spin accumulations. Additionally, the
spin-Coulomb drag reduces the spin accumulation through the spin
diffusion length (see Eq.~(\ref{Ls})). However, in the limit of
$W\ll L_s$,  $\tanh(W/2L_s)$ can be approximated by $W/2L_s$,  and
the spin accumulation at the edges becomes independent of  $L_s$.
In this limit, the influence of SCD on the spin accumulation is
only through the $\lambda\rho_{SD}$ term. Notice that in the limit
of infinite spin-relaxation time ($L_s \to \infty$) the spin
accumulation can be obtained directly from the homogeneous
formulas,  Eqs.~(\ref{Ec}) and (\ref{Es}).

For a two-dimensional
electron gas confined in Al$_{0.1}$Ga$_{0.9}$As quantum well
measured by Sih {\it et al.} \cite{Sih05} with electron and
impurity concentrations $n_i=n_{2D} =2\times 10^{12}$cm$^{-2}$,
mobility $\bar{\mu}$=0.1m$^2$/Vs, $L_s=1\mu$m, $\tau =4\times
10^{-5}$ns, $\tau_{ss} = 0.02$ns, $\alpha\hbar = 4.4 \AA$ and for
the sample with width $W=100\mu$m, we calculate the spin
accumulation to be $-1.5$meV$/|e|$ on the left edge of the sample
(relative to the direction of the electric field) i.e. for
$W/2=-50\mu m$. This means that the non-equilibrium spin-density
points down on the left edge of the sample and up on the right
edge.  It is not clear at present; whether, this sign of the spin
accumulation is consistent or not with the sign of the Kerr
rotation angle measured in the experiments\cite{Sih05,Kato04}.

\section {Summary}
We have developed the microscopic theory of the extrinsic spin
Hall effect taking into account the skew-scattering, side-jump and
Coulomb interaction effects. The total spin conductivity in zero
temperature is a sum of the skew-scattering and side-jump terms.
The spin Coulomb drag is the only temperature dependent term,
causing a quadratic-in-$T$ reduction of the spin Hall
conductivity. Further, we find that the spin Hall conductivity
associated with the side-jump contribution is independent of the
strength of electron-electron interactions, while the part of the
spin conductivity connected with the skew scattering is reduced by
the spin-Coulomb drag for experimentally accessible mobilities.
Moreover, we predict that in paramagnetic systems the spin-Coulomb
drag reduces the spin accumulations in two different ways: (i)
directly through the reduction of the skew-scattering contribution
(ii) indirectly through the reduction of the spin diffusion
length.

\section{Acknowledgements}
This work was supported by NSF Grant No. DMR-0313681. We
gratefully acknowledge valuable discussions with Shufeng Zhang,
Shoucheng Zhang, Masaru Onoda and Roland Winkler. We thank
Professors B. I. Halperin and E. Rashba for their criticism of an
earlier version of this paper and for pointing out to us
Ref.~\cite{Kohn58}.
\end{widetext}

%\bibliography{Extrinsic}

\begin{thebibliography}{44}
\expandafter\ifx\csname
natexlab\endcsname\relax\def\natexlab#1{#1}\fi
\expandafter\ifx\csname bibnamefont\endcsname\relax
  \def\bibnamefont#1{#1}\fi
\expandafter\ifx\csname bibfnamefont\endcsname\relax
  \def\bibfnamefont#1{#1}\fi
\expandafter\ifx\csname citenamefont\endcsname\relax
  \def\citenamefont#1{#1}\fi
\expandafter\ifx\csname url\endcsname\relax
  \def\url#1{\texttt{#1}}\fi
\expandafter\ifx\csname
urlprefix\endcsname\relax\def\urlprefix{URL }\fi
\providecommand{\bibinfo}[2]{#2}
\providecommand{\eprint}[2][]{\url{#2}}

\bibitem[{\citenamefont{Wolf et~al.}(2001)\citenamefont{Wolf, Awschalom,
  Buhrman, Duaghton, von Molnar, Rouke, Chtchelkanova, and Treger}}]{Wolf01}
\bibinfo{author}{\bibfnamefont{S.~A.} \bibnamefont{Wolf}},
  \bibinfo{author}{\bibfnamefont{D.~D.} \bibnamefont{Awschalom}},
  \bibinfo{author}{\bibfnamefont{R.~A.} \bibnamefont{Buhrman}},
  \bibinfo{author}{\bibfnamefont{J.~M.} \bibnamefont{Duaghton}},
  \bibinfo{author}{\bibfnamefont{S.}~\bibnamefont{von Molnar}},
  \bibinfo{author}{\bibfnamefont{M.~L.} \bibnamefont{Rouke}},
  \bibinfo{author}{\bibfnamefont{A.~Y.} \bibnamefont{Chtchelkanova}},
  \bibnamefont{and} \bibinfo{author}{\bibfnamefont{D.~M.}
  \bibnamefont{Treger}}, \bibinfo{journal}{Science}
  \textbf{\bibinfo{volume}{294}}, \bibinfo{pages}{1488} (\bibinfo{year}{2001}).

\bibitem[{\citenamefont{Dyakonov and Perel}(1971{\natexlab{a}})}]{Dyakonov71}
\bibinfo{author}{\bibfnamefont{M.~I.} \bibnamefont{Dyakonov}} \bibnamefont{and}
  \bibinfo{author}{\bibfnamefont{V.~I.} \bibnamefont{Perel}},
  \bibinfo{journal}{Phys. Lett.A} \textbf{\bibinfo{volume}{35}},
  \bibinfo{pages}{459} (\bibinfo{year}{1971}{\natexlab{a}}).

\bibitem[{\citenamefont{Dyakonov and Perel}(1971{\natexlab{b}})}]{Perel}
\bibinfo{author}{\bibfnamefont{M.~I.} \bibnamefont{Dyakonov}} \bibnamefont{and}
  \bibinfo{author}{\bibfnamefont{V.~I.} \bibnamefont{Perel}},
  \bibinfo{journal}{Zh. Eksp. Ter. Fiz.} \textbf{\bibinfo{volume}{13}},
  \bibinfo{pages}{657} (\bibinfo{year}{1971}{\natexlab{b}}).

\bibitem[{\citenamefont{Hirsch}(1999)}]{Hirsch99}
\bibinfo{author}{\bibfnamefont{J.~E.} \bibnamefont{Hirsch}},
  \bibinfo{journal}{Phys. Rev. Lett.} \textbf{\bibinfo{volume}{83}},
  \bibinfo{pages}{1834} (\bibinfo{year}{1999}).

\bibitem[{\citenamefont{Zhang}(2000)}]{Zhang00}
\bibinfo{author}{\bibfnamefont{S.}~\bibnamefont{Zhang}},
  \bibinfo{journal}{Phys. Rev. Lett.} \textbf{\bibinfo{volume}{85}},
  \bibinfo{pages}{393} (\bibinfo{year}{2000}).

\bibitem[{\citenamefont{Murakami et~al.}(2003)\citenamefont{Murakami, Nagaosa,
  and Zhang}}]{Murakami03}
\bibinfo{author}{\bibfnamefont{S.}~\bibnamefont{Murakami}},
  \bibinfo{author}{\bibfnamefont{N.}~\bibnamefont{Nagaosa}}, \bibnamefont{and}
  \bibinfo{author}{\bibfnamefont{S.-C.} \bibnamefont{Zhang}},
  \bibinfo{journal}{Science} \textbf{\bibinfo{volume}{301}},
  \bibinfo{pages}{1348} (\bibinfo{year}{2003}).

\bibitem[{\citenamefont{Sinova et~al.}(2004)\citenamefont{Sinova, Culcer, Niu,
  Sinitsyn, Jungwirth, and MacDonald}}]{Sinova04}
\bibinfo{author}{\bibfnamefont{J.}~\bibnamefont{Sinova}},
  \bibinfo{author}{\bibfnamefont{D.}~\bibnamefont{Culcer}},
  \bibinfo{author}{\bibfnamefont{Q.}~\bibnamefont{Niu}},
  \bibinfo{author}{\bibfnamefont{N.~A.} \bibnamefont{Sinitsyn}},
  \bibinfo{author}{\bibfnamefont{T.}~\bibnamefont{Jungwirth}},
  \bibnamefont{and} \bibinfo{author}{\bibfnamefont{A.~H.}
  \bibnamefont{MacDonald}}, \bibinfo{journal}{Phys. Rev. Lett.}
  \textbf{\bibinfo{volume}{92}}, \bibinfo{pages}{126603}
  (\bibinfo{year}{2004}).

\bibitem[{\citenamefont{Murakami}()}]{Murakami05r}
\bibinfo{author}{\bibfnamefont{S.}~\bibnamefont{Murakami}},
  \eprint{cond-mat/0504353}.

\bibitem[{\citenamefont{Rashba}(2003)}]{Rashba03}
\bibinfo{author}{\bibfnamefont{E.~I.} \bibnamefont{Rashba}},
  \bibinfo{journal}{Phys. Rev. B} \textbf{\bibinfo{volume}{68}},
  \bibinfo{pages}{241315(R)} (\bibinfo{year}{2003}).

\bibitem[{\citenamefont{Kato et~al.}(2004)\citenamefont{Kato, Myers, Gossard,
  and Awschalom}}]{Kato04}
\bibinfo{author}{\bibfnamefont{Y.~K.} \bibnamefont{Kato}},
  \bibinfo{author}{\bibfnamefont{R.~C.} \bibnamefont{Myers}},
  \bibinfo{author}{\bibfnamefont{A.~C.} \bibnamefont{Gossard}},
  \bibnamefont{and} \bibinfo{author}{\bibfnamefont{D.~D.}
  \bibnamefont{Awschalom}}, \bibinfo{journal}{Science}
  \textbf{\bibinfo{volume}{306}}, \bibinfo{pages}{1910} (\bibinfo{year}{2004}).

\bibitem[{\citenamefont{Sih et~al.}(2005)\citenamefont{Sih, Myers, Kato, Lau,
  Gossard, and Awschalom}}]{Sih05}
\bibinfo{author}{\bibfnamefont{V.}~\bibnamefont{Sih}},
  \bibinfo{author}{\bibfnamefont{R.~C.} \bibnamefont{Myers}},
  \bibinfo{author}{\bibfnamefont{Y.~K.} \bibnamefont{Kato}},
  \bibinfo{author}{\bibfnamefont{W.~H.} \bibnamefont{Lau}},
  \bibinfo{author}{\bibfnamefont{A.~C.} \bibnamefont{Gossard}},
  \bibnamefont{and} \bibinfo{author}{\bibfnamefont{D.~D.}
  \bibnamefont{Awschalom}}, \bibinfo{journal}{Nature Physics}
  \textbf{\bibinfo{volume}{1}}, \bibinfo{pages}{31} (\bibinfo{year}{2005}).

\bibitem[{\citenamefont{Wunderlich et~al.}(2005)\citenamefont{Wunderlich,
  Kaestner, Sinova, and Jungwirth}}]{Wunderlich05}
\bibinfo{author}{\bibfnamefont{J.}~\bibnamefont{Wunderlich}},
  \bibinfo{author}{\bibfnamefont{B.}~\bibnamefont{Kaestner}},
  \bibinfo{author}{\bibfnamefont{J.}~\bibnamefont{Sinova}}, \bibnamefont{and}
  \bibinfo{author}{\bibfnamefont{T.}~\bibnamefont{Jungwirth}},
  \bibinfo{journal}{Phys. Rev. Lett.} \textbf{\bibinfo{volume}{94}},
  \bibinfo{pages}{047204} (\bibinfo{year}{2005}).

\bibitem[{\citenamefont{Engel et~al.}()\citenamefont{Engel, Halperin, and
  Rashba}}]{Engel05}
\bibinfo{author}{\bibfnamefont{H.-A.} \bibnamefont{Engel}},
  \bibinfo{author}{\bibfnamefont{B.~I.} \bibnamefont{Halperin}},
  \bibnamefont{and} \bibinfo{author}{\bibfnamefont{E.}~\bibnamefont{Rashba}},
  \eprint{cond-mat/0505535}.

\bibitem[{\citenamefont{Bernevig and Zhang}()}]{Bernevig04}
\bibinfo{author}{\bibfnamefont{B.~A.} \bibnamefont{Bernevig}} \bibnamefont{and}
  \bibinfo{author}{\bibfnamefont{S.-C.} \bibnamefont{Zhang}},
  \eprint{cond-mat/0412550}.

\bibitem[{\citenamefont{Nikolic et~al.}()\citenamefont{Nikolic, Souma, Zarbo,
  and Sinova}}]{Nikolic04}
\bibinfo{author}{\bibfnamefont{B.~K.} \bibnamefont{Nikolic}},
  \bibinfo{author}{\bibfnamefont{S.}~\bibnamefont{Souma}},
  \bibinfo{author}{\bibfnamefont{L.~P.} \bibnamefont{Zarbo}}, \bibnamefont{and}
  \bibinfo{author}{\bibfnamefont{J.}~\bibnamefont{Sinova}},
  \eprint{cond-mat/0412595}.

\bibitem[{\citenamefont{Tso and Sarma}()}]{Sarma05}
\bibinfo{author}{\bibfnamefont{W.-K.} \bibnamefont{Tso}} \bibnamefont{and}
  \bibinfo{author}{\bibfnamefont{S.~D.} \bibnamefont{Sarma}},
  \eprint{cond-mat/0502426}.

\bibitem[{\citenamefont{Tse et~al.}()\citenamefont{Tse, Fabian, Zutic, and
  Sarma}}]{Sarma05a}
\bibinfo{author}{\bibfnamefont{W.-K.} \bibnamefont{Tse}},
  \bibinfo{author}{\bibfnamefont{J.}~\bibnamefont{Fabian}},
  \bibinfo{author}{\bibfnamefont{I.}~\bibnamefont{Zutic}}, \bibnamefont{and}
  \bibinfo{author}{\bibfnamefont{S.~D.} \bibnamefont{Sarma}},
  \eprint{cond-mat/0508076}.

\bibitem[{\citenamefont{Nozi\'eres and Lewiner}(1973)}]{Nozieres}
\bibinfo{author}{\bibfnamefont{P.}~\bibnamefont{Nozi\'eres}} \bibnamefont{and}
  \bibinfo{author}{\bibfnamefont{C.}~\bibnamefont{Lewiner}},
  \bibinfo{journal}{J. Phys. (Paris)} \textbf{\bibinfo{volume}{34}},
  \bibinfo{pages}{901} (\bibinfo{year}{1973}).

\bibitem[{\citenamefont{Smit}(1955)}]{Smit55}
\bibinfo{author}{\bibfnamefont{J.}~\bibnamefont{Smit}},
  \bibinfo{journal}{Physica} \textbf{\bibinfo{volume}{21}},
  \bibinfo{pages}{877} (\bibinfo{year}{1955}).

\bibitem[{\citenamefont{Smit}(1958)}]{Smit58}
\bibinfo{author}{\bibfnamefont{J.}~\bibnamefont{Smit}},
  \bibinfo{journal}{Physica} \textbf{\bibinfo{volume}{24}}, \bibinfo{pages}{39}
  (\bibinfo{year}{1958}).

\bibitem[{\citenamefont{Mott and Massey}(1964)}]{Mott}
\bibinfo{author}{\bibfnamefont{N.~F.} \bibnamefont{Mott}} \bibnamefont{and}
  \bibinfo{author}{\bibfnamefont{H.~S.~W.} \bibnamefont{Massey}},
  \emph{\bibinfo{title}{The Theory of Atomic Collisions}}
  (\bibinfo{publisher}{Oxford University Press}, \bibinfo{year}{1964}).

\bibitem[{\citenamefont{Berger}(1970)}]{Berger70a}
\bibinfo{author}{\bibfnamefont{L.}~\bibnamefont{Berger}},
  \bibinfo{journal}{Phys. Rev. B} \textbf{\bibinfo{volume}{2}},
  \bibinfo{pages}{4559} (\bibinfo{year}{1970}).

\bibitem[{\citenamefont{Berger}(1972)}]{Berger70b}
\bibinfo{author}{\bibfnamefont{L.}~\bibnamefont{Berger}},
  \bibinfo{journal}{Phys. Rev. B} \textbf{\bibinfo{volume}{5}},
  \bibinfo{pages}{1862} (\bibinfo{year}{1972}).

\bibitem[{\citenamefont{Lyo and Holstein}(1972)}]{Lyo72}
\bibinfo{author}{\bibfnamefont{S.~K.} \bibnamefont{Lyo}} \bibnamefont{and}
  \bibinfo{author}{\bibfnamefont{T.}~\bibnamefont{Holstein}},
  \bibinfo{journal}{Phys. Rev. Lett.} \textbf{\bibinfo{volume}{29}},
  \bibinfo{pages}{423} (\bibinfo{year}{1972}).

\bibitem[{\citenamefont{Karplus and Luttinger}(1954)}]{Luttinger54}
\bibinfo{author}{\bibfnamefont{R.}~\bibnamefont{Karplus}} \bibnamefont{and}
  \bibinfo{author}{\bibfnamefont{J.~M.} \bibnamefont{Luttinger}},
  \bibinfo{journal}{Phys. Rev. B} \textbf{\bibinfo{volume}{95}},
  \bibinfo{pages}{1154} (\bibinfo{year}{1954}).

\bibitem[{\citenamefont{Dugaev et~al.}(2001)\citenamefont{Dugaev, Cr\'epieux,
  and Bruno}}]{Dugaev01}
\bibinfo{author}{\bibfnamefont{V.~K.} \bibnamefont{Dugaev}},
  \bibinfo{author}{\bibfnamefont{A.}~\bibnamefont{Cr\'epieux}},
  \bibnamefont{and} \bibinfo{author}{\bibfnamefont{P.}~\bibnamefont{Bruno}},
  \bibinfo{journal}{Phys. Rev. B} \textbf{\bibinfo{volume}{64}},
  \bibinfo{pages}{104411} (\bibinfo{year}{2001}).

\bibitem[{\citenamefont{Cr\'epieux and Bruno}(2001)}]{Bruno01}
\bibinfo{author}{\bibfnamefont{A.}~\bibnamefont{Cr\'epieux}} \bibnamefont{and}
  \bibinfo{author}{\bibfnamefont{P.}~\bibnamefont{Bruno}},
  \bibinfo{journal}{Phys. Rev. B} \textbf{\bibinfo{volume}{64}},
  \bibinfo{pages}{014416} (\bibinfo{year}{2001}).

\bibitem[{\citenamefont{Jungwirth et~al.}(2002)\citenamefont{Jungwirth, Niu,
  and MacDonald}}]{Jungwirth02}
\bibinfo{author}{\bibfnamefont{T.}~\bibnamefont{Jungwirth}},
  \bibinfo{author}{\bibfnamefont{Q.}~\bibnamefont{Niu}}, \bibnamefont{and}
  \bibinfo{author}{\bibfnamefont{A.~H.} \bibnamefont{MacDonald}},
  \bibinfo{journal}{Phys. Rev. Lett.} \textbf{\bibinfo{volume}{88}},
  \bibinfo{pages}{207208} (\bibinfo{year}{2002}).

\bibitem[{\citenamefont{Onoda and Nagaosa}(2003)}]{Onoda02}
\bibinfo{author}{\bibfnamefont{M.}~\bibnamefont{Onoda}} \bibnamefont{and}
  \bibinfo{author}{\bibfnamefont{N.}~\bibnamefont{Nagaosa}},
  \bibinfo{journal}{Phys. Rev. Lett.} \textbf{\bibinfo{volume}{90}},
  \bibinfo{pages}{206601} (\bibinfo{year}{2003}).

\bibitem[{\citenamefont{Dugaev et~al.}()\citenamefont{Dugaev, Bruno,
  Taillefumier, Canals, and Lacroix}}]{Bruno05}
\bibinfo{author}{\bibfnamefont{V.~K.} \bibnamefont{Dugaev}},
  \bibinfo{author}{\bibfnamefont{P.}~\bibnamefont{Bruno}},
  \bibinfo{author}{\bibfnamefont{M.}~\bibnamefont{Taillefumier}},
  \bibinfo{author}{\bibfnamefont{B.}~\bibnamefont{Canals}}, \bibnamefont{and}
  \bibinfo{author}{\bibfnamefont{C.}~\bibnamefont{Lacroix}},
  \eprint{cond-mat/0502386}.

\bibitem[{\citenamefont{Sinitsyn et~al.}()\citenamefont{Sinitsyn, Niu, Sinova,
  and Nomura}}]{Sinitsyn05}
\bibinfo{author}{\bibfnamefont{N.~A.} \bibnamefont{Sinitsyn}},
  \bibinfo{author}{\bibfnamefont{Q.}~\bibnamefont{Niu}},
  \bibinfo{author}{\bibfnamefont{J.}~\bibnamefont{Sinova}}, \bibnamefont{and}
  \bibinfo{author}{\bibfnamefont{K.}~\bibnamefont{Nomura}},
  \eprint{cond-mat/0502426}.

\bibitem[{\citenamefont{D'Amico and Vignale}(2000)}]{Amico00}
\bibinfo{author}{\bibfnamefont{I.}~\bibnamefont{D'Amico}} \bibnamefont{and}
  \bibinfo{author}{\bibfnamefont{G.}~\bibnamefont{Vignale}},
  \bibinfo{journal}{Phys. Rev. B} \textbf{\bibinfo{volume}{62}},
  \bibinfo{pages}{4853} (\bibinfo{year}{2000}).

\bibitem[{\citenamefont{Flensberg et~al.}(2001)\citenamefont{Flensberg, Jensen,
  and Mortensen}}]{Flensberg01}
\bibinfo{author}{\bibfnamefont{K.}~\bibnamefont{Flensberg}},
  \bibinfo{author}{\bibfnamefont{T.~S.} \bibnamefont{Jensen}},
  \bibnamefont{and} \bibinfo{author}{\bibfnamefont{N.~A.}
  \bibnamefont{Mortensen}}, \bibinfo{journal}{Phys. Rev. B}
  \textbf{\bibinfo{volume}{64}}, \bibinfo{pages}{245308}
  (\bibinfo{year}{2001}).

\bibitem[{\citenamefont{D'Amico and Vignale}(2002)}]{Amico02}
\bibinfo{author}{\bibfnamefont{I.}~\bibnamefont{D'Amico}} \bibnamefont{and}
  \bibinfo{author}{\bibfnamefont{G.}~\bibnamefont{Vignale}},
  \bibinfo{journal}{Phys. Rev. B} \textbf{\bibinfo{volume}{65}},
  \bibinfo{pages}{85109} (\bibinfo{year}{2002}).

\bibitem[{\citenamefont{Weber et~al.}(2005)\citenamefont{Weber, Gedik, Moore,
  Orenstein, Stephens, and Awschalom}}]{Orenstein05}
\bibinfo{author}{\bibfnamefont{C.~P.} \bibnamefont{Weber}},
  \bibinfo{author}{\bibfnamefont{N.}~\bibnamefont{Gedik}},
  \bibinfo{author}{\bibfnamefont{J.~E.} \bibnamefont{Moore}},
  \bibinfo{author}{\bibfnamefont{J.}~\bibnamefont{Orenstein}},
  \bibinfo{author}{\bibfnamefont{J.}~\bibnamefont{Stephens}}, \bibnamefont{and}
  \bibinfo{author}{\bibfnamefont{D.~D.} \bibnamefont{Awschalom}},
  \bibinfo{journal}{Bull. Am. Phys. Soc.} \textbf{\bibinfo{volume}{50}},
  \bibinfo{pages}{1320} (\bibinfo{year}{2005}), \bibinfo{note}{(Abstracts W
  10.1 and W 10.2)}.

\bibitem[{\citenamefont{Winkler}(2003)}]{Winkler2003}
\bibinfo{author}{\bibfnamefont{R.}~\bibnamefont{Winkler}},
  \emph{\bibinfo{title}{Spin-orbit effects in two-dimensional electron and hole
  systems}} (\bibinfo{publisher}{Springer}, \bibinfo{year}{2003}).

\bibitem[{\citenamefont{Kohn and Luttinger}(1957)}]{Kohn58}
\bibinfo{author}{\bibfnamefont{W.}~\bibnamefont{Kohn}} \bibnamefont{and}
  \bibinfo{author}{\bibfnamefont{J.~M.} \bibnamefont{Luttinger}},
  \bibinfo{journal}{Phys. Rev.} \textbf{\bibinfo{volume}{108}},
  \bibinfo{pages}{590} (\bibinfo{year}{1957}).

\bibitem[{\citenamefont{Landau and Lifshitz}(1964)}]{Landau3}
\bibinfo{author}{\bibfnamefont{L.~D.} \bibnamefont{Landau}} \bibnamefont{and}
  \bibinfo{author}{\bibfnamefont{E.~M.} \bibnamefont{Lifshitz}},
  \emph{\bibinfo{title}{Course of Theoretical Physics, Vol. III.}}
  (\bibinfo{publisher}{Butterworth-Heinemann}, \bibinfo{address}{Oxford},
  \bibinfo{year}{1964}).

\bibitem[{\citenamefont{D'Amico and Vignale}(2003)}]{Amico03}
\bibinfo{author}{\bibfnamefont{I.}~\bibnamefont{D'Amico}} \bibnamefont{and}
  \bibinfo{author}{\bibfnamefont{G.}~\bibnamefont{Vignale}},
  \bibinfo{journal}{Phys. Rev. B} \textbf{\bibinfo{volume}{68}},
  \bibinfo{pages}{045307} (\bibinfo{year}{2003}).

\bibitem[{\citenamefont{Mishchenko et~al.}(2004)\citenamefont{Mishchenko,
  Shytov, and Halperin}}]{Halperin04}
\bibinfo{author}{\bibfnamefont{E.~G.} \bibnamefont{Mishchenko}},
  \bibinfo{author}{\bibfnamefont{A.~V.} \bibnamefont{Shytov}},
  \bibnamefont{and} \bibinfo{author}{\bibfnamefont{B.~I.}
  \bibnamefont{Halperin}}, \bibinfo{journal}{Phys. Rev. Lett.}
  \textbf{\bibinfo{volume}{93}}, \bibinfo{pages}{226602}
  (\bibinfo{year}{2004}).

\bibitem[{\citenamefont{Khaetskii}()}]{Khaetskii04}
\bibinfo{author}{\bibfnamefont{A.}~\bibnamefont{Khaetskii}},
  \eprint{cond-mat/0408136}.

\bibitem[{\citenamefont{Raimondi and Schwab}()}]{Schwab04}
\bibinfo{author}{\bibfnamefont{R.}~\bibnamefont{Raimondi}} \bibnamefont{and}
  \bibinfo{author}{\bibfnamefont{P.}~\bibnamefont{Schwab}},
  \eprint{cond-mat/0408233}.

\bibitem[{\citenamefont{Hankiewicz et~al.}()\citenamefont{Hankiewicz, Vignale,
  and Flatt\'e}}]{Hankiewicz_unpublished}
\bibinfo{author}{\bibfnamefont{E.~M.} \bibnamefont{Hankiewicz}},
  \bibinfo{author}{\bibfnamefont{G.}~\bibnamefont{Vignale}}, \bibnamefont{and}
  \bibinfo{author}{\bibfnamefont{M.}~\bibnamefont{Flatt\'e}},
  \eprint{unpublished}.

\bibitem[{\citenamefont{Valet and Fert}(1993)}]{Fert93}
\bibinfo{author}{\bibfnamefont{T.}~\bibnamefont{Valet}} \bibnamefont{and}
  \bibinfo{author}{\bibfnamefont{A.}~\bibnamefont{Fert}},
  \bibinfo{journal}{Phys. Rev. B} \textbf{\bibinfo{volume}{48}},
  \bibinfo{pages}{7099} (\bibinfo{year}{1993}).

\end{thebibliography}

\begin{widetext}
\section{Appendix- An exactly solvable model for skew-scattering}
We present the calculations of the scattering rates
   for a circular well potential of the form:
\begin{equation}\label{scatter_V}
V(r)= V_0\theta(a-r)+ \bar{\alpha}aL_zS_z\delta(r-a)V_0 ~,
\end{equation}
which is attractive for $V_0<0$ and repulsive for $V_0>0$. The
second term on the right is the spin-orbit interaction and
$\bar{\alpha} = \alpha \hbar/a^2$ where $\alpha$ is an effective
spin-orbit coupling found in 8-band Kane model and $a$ is the
impurity radius. Since the orbital angular momentum $L_z=l$ and
the spin angular momentum $S_z=\sigma$ are conserved we can
separate the wave function into radial and orbital parts:
\begin{equation}
\Psi_{kl\sigma}(r,\theta) =R_{kl\sigma}(r)e^{il\theta}
\end{equation}
and the corresponding Schr\"odinger equation has the form:
\begin{equation}
-\frac{\hbar^2}{2m}\left(R''_{kl\sigma}+\frac{1}{r}R'_{kl\sigma}\right)
+\frac{\hbar^2l^2}{2mr^2}R_{kl\sigma}+V(r)R_{kl\sigma}(r)
=ER_{kl\sigma}(r)
\end{equation}
where $E=\hbar^2k^2/2m$ and the prime denotes a derivative with
respect to $r$. We now express lengths and wave vectors in units
$a$ and $a^{-1}$ respectively, so $r$ should be understood as
$r/a$, $k$ as $ka$, and, of course, $a=1$ in these units. The
dimensionless Schr\"odinger equation is
\begin{equation}\label{diffeq.1}
R''_{kl\sigma}+\frac{1}{r}R'_{kl\sigma}
+\left(k^2-v_0-\frac{l^2}{r^2}\right)R_{kl\sigma} = 0, ~~~~r<1
\end{equation}
and
\begin{equation} \label{diffeq.2}
R''_{kl\sigma}+\frac{1}{r}R'_{kl\sigma}
+\left(k^2-\frac{l^2}{r^2}\right)R_{kl\sigma} = 0,~~~~r>1~,
\end{equation}
where $v_0 = \frac{2mV_0a^2}{\hbar^2}$ is the dimensionless
parameter which measures the height of the impurity potential
barrier. The regular solution of this equation for $r<1$ is
\begin{equation}\label{radial1}
R_{kl\sigma}(r)=J_{|l|}(\nu r), ~~~~r<1~,
\end{equation}
where we have defined $\nu =\sqrt{k^2-v_0}$. On the other hand,
the solution for $r>1$ can be written as a superposition of the
two independent solutions of the differential
equation~(\ref{diffeq.2}):
\begin{equation}\label{radial2}
R_{kl\sigma}(r)=e^{i\delta_{l\sigma}}
[\cos\delta_{l\sigma}J_{|l|}(kr)-\sin\delta_{l\sigma}Y_{|l|}(kr)],
r>1
\end{equation}
The matching conditions on the wave function and its derivative
lead to the following equations:
\begin{equation}
R_{kl\sigma}(1^{+})=R_{kl\sigma}(1^{-})=R_{kl\sigma}(1)
\end{equation}
\begin{equation}
R'_{kl\sigma}(1^{+})-R'_{kl\sigma}(1^{-})= \bar{\alpha} l\sigma
R_{kl\sigma}(1)v_0
\end{equation}
Substitution of Eqs.~(\ref{radial1}) and (\ref{radial2}) to the
matching conditions yields
\begin{equation}
\frac{k \cos\delta_{l\sigma}J'_{|l|}(k)-k
\sin\delta_{l\sigma}Y'_{|l|}(k)}
{\cos\delta_{l\sigma}J_{|l|}(k)-\sin\delta_{l\sigma}Y_{|l|}(k)}=
\nu\frac{J'_{|l|}(\nu)}{J_{|l|}(\nu)}+\bar{\alpha}l\sigma v_0~,
\end{equation}
from which one gets the following equation for the phase shifts
$\delta_{l\sigma}$:
\begin{equation}\label{phase.shifts}
\cot\delta_{l\sigma} = \frac{k
Y'_{|l|}(k)-\beta_{l\sigma}Y_{|l|}(k)}{kJ'_{|l|}(k)-\beta_{l\sigma}J_{|l|}(k)}
\end{equation}
where
$\beta_{l\sigma}=\nu\frac{J'_{|l|}(\nu)}{J_{|l|}(\nu)}+\bar{\alpha}l\sigma
v_0$, $J_{|l|}(k)$ and $Y_{|l|}(k)$ are the Bessel functions of
the first and second kind.
  The wave function at large distance
from the scattering center can be written as:
\begin{equation}
\psi_{kl\sigma}(r,\theta) \stackrel{r\rightarrow\infty}{\sim}\sqrt{\frac{2}{\pi
kr}}\cos{(kr-\frac{|l|\pi}{2}-\frac{\pi}{4}+\delta_{l\sigma})}e^{il\theta}e^{i\delta_{l\sigma}}
=\psi^{0}_{kl\sigma}(r,\theta)
+\frac{e^{2i\delta_{l\sigma}}-1}{\sqrt{2\pi
kr}}e^{i(kr-|l|\pi/2-\pi/4)}e^{il\theta}
\end{equation}
where $\psi_{kl\sigma}^{0}(r,\theta)=\sqrt{\frac{2}{\pi
kr}}\cos{(kr-\frac{|l|\pi}{2}-\frac{\pi}{4})}e^{il\theta}$ is the
free wave function in the channel of angular momentum $l$. The
scattering amplitude $f_\sigma(k,\theta)$ is the factor multiplying
the outgoing wave
$\frac{e^{ikr}}{\sqrt{r}}$ in the above equation:
\begin{equation}
f_{\sigma}(k,\theta) = \sum_{l=-\infty}^{\infty}
\frac{e^{2i\delta_{l\sigma}}-1}{\sqrt{2\pi
k}}e^{-i(|l|\pi/2+\pi/4)}e^{il\theta}
\end{equation}
The differential cross section is accordingly given by:
\begin{equation}\label{differential.cross.section}
{\left(
\frac{d\sigma_c}{d\theta}\right)}_{\sigma}=|f_{\sigma}(k,\theta)|^2
=\frac{1}{2\pi
k}\sum_{l,l'}(e^{2i\delta_{l\sigma}}-1)(e^{-2i\delta_{l'\sigma}}-1)e^{-i\pi/2(|l|-|l'|)}e^{i(l-l')\theta}
\end{equation}

Finally we notice that the total scattering rate is related
to the differential scattering cross section for a single impurity as follows:
\begin{equation}\label{scattering.probability.2}
W(k,\theta)= W^s(k,\theta)+\sigma W^a(k,\theta)  \sin \theta =n_i
\frac{4\pi^2 \hbar^3 k}{m^2{\cal A}} \frac{d\sigma_c}{d\theta}~,
\end{equation}
where $n_i=N_i/{\cal A}$ is the areal density of impurities.
Combining this with Eq.~(\ref{differential.cross.section}) we find
\begin{equation}\label{scattering.probability.a1}
W(k,\theta) = n_i \frac{2\pi^2 \hbar^3}{m^2{\cal
A}}\sum_{l,l'}(e^{2i\delta_{l\sigma}}-1)(e^{-2i\delta_{l'\sigma}}-1)i^{|l'|-|l|}e^{i(l-l')\theta},
\end{equation}
To identify $W^{s}$ and $W^{a}$ we separate
Eq.~(\ref{scattering.probability.a1}) into even and odd components
with respect to the scattering angle $\theta$, which can be easily
done using the identity $e^{i(l-l')\theta}= \cos[(l-l')\theta] +i
\sin[(l-l')\theta]$. Then
\begin{equation}\label{scattering.probability.plus1}
W^s(k,\theta)= n_i \frac{2\pi^2 \hbar^3}{m^2{\cal
A}}\sum_{l,l'}(e^{2i\delta_{l\sigma}}-1)(e^{-2i\delta_{l'\sigma}}-1)i^{|l'|-|l|}\cos[(l-l')\theta]~
\end{equation}
and
\begin{equation}\label{scattering.probability.minus1}
W^a(k,\theta)= \sigma n_i\frac{2\pi^2 \hbar^3}{m^2{\cal A}
\sin\theta}\sum_{l,l'}(e^{2i\delta_{l\sigma}}-1)(e^{-2i\delta_{l'\sigma}}-1)i^{|l'|-|l|+1}\sin[(l-l')\theta],
\end{equation}
Making use of the identities $e^{\pm 2i\delta_{l\sigma}}-1=\frac{\pm
2i}{\cot\delta_{l\sigma}\mp i}$ we rewrite the scattering rates as
\begin{equation}\label{scattering.probability.plus2}
W^s(k,\theta)= n_i\frac{8\pi^2 \hbar^3}{m^2{\cal
A}}\sum_{l,l'}\frac{i^{|l'|-|l|}\cos[(l-l')\theta]}{(\cot\delta_{l\sigma}-i)
(\cot\delta_{l'\sigma}+i)}
\end{equation}
and
\begin{equation}\label{scattering.probability.minus2}
W^a(k,\theta)= \sigma n_i\frac{8\pi^2 \hbar^3}{m^2{\cal
A}}\sum_{l,l'}\frac{i^{|l'|-|l|+1}\sin[(l-l')\theta]}{(\cot\delta_{l\sigma}-i)
(\cot\delta_{l'\sigma}+i) \sin\theta}~.
\end{equation}
where the phase shifts are completely determined by
Eq.~(\ref{phase.shifts}). Notice that the sums over $l$ and $l'$
in Eq.~(\ref{scattering.probability.plus2}) and
Eq.~(\ref{scattering.probability.minus2}) run from $-\infty$ to
$\infty$ and the phase shifts have the symmetries
$\delta_{-l,-\sigma}(\alpha)=\delta_{l,\sigma}(\alpha)$ and
$\delta_{-l,\sigma}(-\alpha)=\delta_{l,\sigma}(\alpha)$, which
implies that $W^s(k,\theta)$ and $W^a(k,\theta)$ are invariant
under spin reversal $\sigma \rightarrow -\sigma$ and
$W^a(k,\theta)$ changes sign with a change of sign of $\alpha$, as
expected. The integral over $\theta$ (Eq.~\ref{tau.regular.2})
eliminates the majority of the terms from the sum over $l$ and
$l'$ and the only non-zero terms for $W^s(k,\theta)$ are with
$l=l'$ and $l=l'\pm 1$. This gives
\begin{equation}\label{scattering.probability.plus3}
\int_0^{2\pi} d \theta~ W^s(k,\theta)(1-\cos\theta)=
n_i\frac{8\pi^3 \hbar^3}{m^2{\cal A}}\left\{\sum_{l}
\frac{2}{(\cot\delta_{l\sigma}^2+1)}-\sum_{l=l'\pm 1}
\frac{i^{|l'|-|l|}}{(\cot\delta_{l\sigma}-i)
(\cot\delta_{l'\sigma}+i)} ~\right\}.
\end{equation}
One can see that $W^{s}(k,\theta)$ is modified by spin-orbit
interactions, however it has a non-zero value even if spin-orbit
interactions are absent. We checked numerically that the sum over
$l$ in (Eq.~\ref{scattering.probability.plus3}) is convergent
after taking into account a few first terms. For the
skew-scattering rate $W^{a}(k,\theta)$, the integral over $\theta$
(Eq.~\ref{tau.anomalous.2}) have non-zero terms only if $l=l'\pm
1$. This gives:
\begin{equation}\label{scattering.probability.minus3}
\int_0^{2\pi} d \theta~ W^a(k,\theta)\sin^2 \theta = \sigma
n_i\frac{8\pi^3 \hbar^3}{m^2{\cal A}}\sum_{l=l'\pm
1}\frac{i^{|l'|-|l|+1}}{(\cot\delta_{l\sigma}-i)
(\cot\delta_{l'\sigma}+i)}~.
\end{equation}
For very small $ka$ the only relevant terms are $l'=0$, $|l|=1$
and $|l'|=1$, $l=0$ which yields to:
\begin{eqnarray}\label{scattering.probability.minus4}
\int_0^{2\pi} d \theta~ W^a(k,\theta)\sin^2 \theta \simeq\sigma
n_i\frac{8\pi^3 \hbar^3}{m^2{\cal
A}}\left[\frac{1}{(\cot\delta_{1\sigma}-i)(\cot\delta_{0\sigma}+i)}
-\frac{1}{(\cot\delta_{0\sigma}-i)(\cot\delta_{-1\sigma}+i)}+\right.\nonumber\\
\left.\frac{1}{(\cot\delta_{0\sigma}-i)(\cot\delta_{1\sigma}+i)}
-\frac{1}{(\cot\delta_{-1\sigma}-i)(\cot\delta_{0\sigma}+i)}\right]=\nonumber\\
\sigma n_i\frac{16\pi^2 \hbar^3}{m^2{\cal
A}}\left[Re\frac{1}{(\cot\delta_{1\sigma}-i)(\cot\delta_{0\sigma}+i)}
-Re\frac{1}{(\cot\delta_{-1\sigma}-i)(\cot\delta_{0\sigma}+i)}\right]=\nonumber\\
=\sigma n_i\frac{16\pi^2 \hbar^3}{m^2{\cal
A}}\left[\frac{1+\cot\delta_{0\sigma}\cot\delta_{1\sigma}}{(\cot^2\delta_{1\sigma}+1)(\cot^2\delta_{0\sigma}+1)}
-\frac{1+\cot\delta_{0\sigma}\cot\delta_{-1\sigma}}{(\cot^2\delta_{-1\sigma}+1)(\cot^2\delta_{0\sigma}+1)}\right]~,
\end{eqnarray}
independent of $\theta$.  Notice that $\int_0^{2\pi} d \theta~
W^a(k,\theta)\sin^2 \theta$ vanishes in the absence of spin-orbit
interactions, since all the phase shifts are independent of
$\sigma$ in that case.
\end{widetext}

\end{document}